\documentclass[11pt]{article}
\pdfoutput=1

\usepackage{amssymb}
\usepackage{amsmath}
\usepackage{bbold}
\usepackage{color}
\usepackage{colordvi}
\usepackage{colortbl}
\usepackage{fancybox}
\usepackage[footnotesize]{caption2}
\usepackage{graphicx}
\usepackage[center,footnotesize,hang]{subfigure}
\usepackage{bbm}
\usepackage{url}
\usepackage{multirow}
\usepackage{array}
\usepackage{arydshln}
\usepackage{pifont}
\usepackage{mathrsfs}
\usepackage{fancybox}
\usepackage{cite}
\usepackage[colorlinks]{hyperref}
\usepackage{enumitem}
\usepackage{float}
\usepackage{longtable}
\newcommand{\PreserveBackslash}[1]{\let\temp=\\#1\let\\=\temp}
\newcolumntype{C}[1]{>{\PreserveBackslash\centering}p{#1}}
\newcolumntype{R}[1]{>{\PreserveBackslash\raggedleft}p{#1}}
\newcolumntype{L}[1]{>{\PreserveBackslash\raggedright}p{#1}}
\allowdisplaybreaks \allowdisplaybreaks[2]

\definecolor{lightred}{rgb}{1,0.4,0.4}
\definecolor{lightgreen}{rgb}{0.4,1,0.4}
\definecolor{LightCyan}{rgb}{0.88,1,1}

\begin{document}

\title{
\begin{flushright}
\hfill\mbox{\small USTC-ICTS-19-15} \\[5mm]
\begin{minipage}{0.2\linewidth}
\normalsize
\end{minipage}
\end{flushright}
{\Large \bf A New Littlest Seesaw Model
\\[2mm]}}

\date{}

\author{
Ping-Tao~Chen$^{1}$\footnote{E-mail: {\tt
chenpt@mail.ustc.edu.cn}},  \
Gui-Jun~Ding$^{1}$\footnote{E-mail: {\tt
dinggj@ustc.edu.cn}},  \
Stephen~F.~King$^{2}$\footnote{E-mail: {\tt king@soton.ac.uk}}, \
Cai-Chang Li$^{1,3}$\footnote{E-mail: {\tt
lcc0915@mail.ustc.edu.cn}}  \
\\*[20pt]
\centerline{
\begin{minipage}{\linewidth}
\begin{center}
$^1${\it \small
Interdisciplinary Center for Theoretical Study and  Department of Modern Physics,\\
University of Science and Technology of China, Hefei, Anhui 230026, China}\\[2mm]
$^2${\it \small
Physics and Astronomy,
University of Southampton,
Southampton, SO17 1BJ, U.K.}\\[2mm]
$^3${\it \small
School of Physics, 
Northwest University, 
Xi'an 710127, China}\\
\end{center}
\end{minipage}}
\\[10mm]}
\maketitle
\thispagestyle{empty}

\begin{abstract}
\noindent

We propose and discuss a new Littlest Seesaw model,  realized in the tri-direct CP approach, in which the couplings of the two right-handed neutrinos to the lepton doublets are proportional to $(0,-1,1)$ and $(1,5/2,-1/2)$ respectively with the relative phase $\eta=-\pi/2$. This model can give an excellent description of lepton flavour mixing, including an atmospheric neutrino mixing angle in the second octant,
in terms of only two input parameters. We show that the observed baryon asymmetry can be generated for the lightest right-handed neutrino mass $M_{1}=1.176\times 10^{11}$ GeV in SM and $M_{1}=3.992\times 10^{10}$ GeV in MSSM with $\tan\beta=5$. We construct an explicit Littlest Seesaw model based on the flavour symmetry $S_4\times Z_5\times Z_8$ in which the desired alignments and the phase $\eta=-\pi/2$ are achieved.

\end{abstract}
\newpage

\section{Introduction }

The Standard Model (SM) has been well established by the discovery of the Higgs boson. However, the discovery of neutrino oscillations implies that neutrinos have masses and there are flavour mixing in lepton sector. Non-zero neutrino masses open up a window to the new physics beyond SM. However, the origin of neutrino mass generation and the flavour mixings in quark and lepton sectors are still unknown~\cite{King:2013eh,King:2015aea}. In order to elegantly generate the tiny neutrino mass, the most appealing theory seems to be type I seesaw mechanism involving heavy right-handed Majorana neutrinos~\cite{Minkowski:1977sc,Mohapatra:1979ia,Schechter:1980gr}.

The type I seesaw mechanism can qualitatively explain the smallness of neutrino masses through the heavy right-handed neutrinos. However, if one doesn't make other assumptions, the seesaw model with three right-handed neutrinos (RHN) contains too many parameters to make any particular predictions for neutrino mass and mixing. As we know, the idea of sequential dominance (SD)~\cite{King:1998jw,King:1999cm} of right-handed neutrinos  is an effective method to produce the mass hierarchy between the two mass squared differences $\Delta m^2_{21}$ and $\Delta m^2_{31}$~\cite{Esteban:2018azc}, it requires that the mass spectrum of heavy Majorana neutrinos is strongly hierarchical, i.e. $M_\text{atm}\ll M_\text{sol}\ll M_\text{dec}$.
It arises from the proposal that a dominant heavy right-handed (RH) neutrino is mainly
responsible for the atmospheric neutrino mass, a heavier subdominant RH
neutrino for the solar neutrino mass, and a possible third largely decoupled RH
neutrino for the lightest neutrino mass. It leads to an effective two right-handed neutrino (2RHN) model ~\cite{King:1999mb,Frampton:2002qc}. This simple idea leads to equally simple predictions which makes the scheme falsifiable. Indeed, the litmus test of such SD is a very light (or massless) neutrino. These predictions will be tested soon. In order to further increase predictive power of the minimal seesaw mechanism, various proposals have been suggested, such as postulating
one~\cite{King:2002nf} or two~\cite{Frampton:2002qc} texture zeros in the neutrino Yukawa coupling. The models with two texture zero are excluded by the present data for normal ordering neutrino masses~\cite{Guo:2006qa,Harigaya:2012bw,Zhang:2015tea}.

A very predictive minimal seesaw model with one texture zero is the so-called CSD($n$) model ~\cite{King:2005bj,Antusch:2011ic,King:2013iva,King:2015dvf,King:2016yvg,Ballett:2016yod,King:2018fqh,King:2013xba,King:2013hoa,Bjorkeroth:2014vha}, where the parameter $n$ was usually assumed to be a positive integer. The CSD($n$) scheme assumes that the two columns of the Dirac neutrino mass matrix are proportional to $(0,1, -1)$ and $(1, n, 2-n)$ respectively in the RHN diagonal basis. As a consequence, the lepton mixing matrix is predicted to be TM1 pattern, the neutrino masses are normal ordering and the lightest neutrino is massless with $m_1=0$. At present only the CSD($3$) (also called Littlest Seesaw model)~\cite{King:2013iva,King:2015dvf,King:2016yvg,Ballett:2016yod,King:2018fqh} and CSD($4$) models~\cite{King:2013xba,King:2013hoa} can give rise to phenomenologically viable predictions for lepton mixing parameters and the two neutrino
mass squared differences $\Delta m^2_{21}$ and $\Delta m^2_{31}$.

It has been shown that CSD($n$) can be enforced by a residual symmetry of $S_4$
\cite{King:2016yvg} in the semi-direct approach where different residual flavour symmetries $G_{l}=Z^T_3$
and $G_{\nu}=Z_2^{SU}$ are assumed in the charged lepton
and neutrino sectors. However, it was not possible to identify any residual CP symmetry
for CSD($n$) in the semi-direct approach. This means that the parameter $n$ of CSD($n$), which is usually assumed to be integer valued, could in fact be a complex number in general. In order to preserve the predictions of CSD($n$), we would like to fix the parameter $n$ to be real (although not necessarily an integer). This suggests that we should seek to somehow use residual CP symmetry, even though it is not possible within the semi-direct approach.

In the past years, discrete flavour symmetry has been combined with generalized CP symmetry to provide a powerful framework to explain the lepton mixing angles and predict leptonic CP violation phases~\cite{Feruglio:2012cw,Holthausen:2012dk,Ding:2013hpa,Ding:2013bpa,Li:2013jya,Ding:2013nsa,
Ding:2014ssa,Ding:2014hva,Li:2014eia,Ding:2014ora,Chen:2014wxa,Everett:2015oka,Branco:2015hea,Li:2015jxa,DiIura:2015kfa,Ballett:2015wia,Branco:2015gna,Chen:2015nha,Ding:2015rwa,
Chen:2015siy,Li:2016ppt,Chen:2016ptr,Yao:2016zev,Li:2016nap,Lu:2016jit,Everett:2016jsk,Li:2017zmk,Li:2017abz,Lu:2018oxc,Lu:2019gqp,Chen:2018lsv,Hagedorn:2016lva,Delgadillo:2018tza}.
Furthermore, a simultaneous description of quark and lepton flavour mixing and CP violation can be achieved through spontaneous breaking of a discrete family symmetry and CP symmetry~\cite{Li:2017abz,Lu:2018oxc,Lu:2019gqp}. Since the generalized CP symmetry may play a critical role in understanding the flavour puzzle of SM, recently we extended the widely studied direct model of discrete flavour symmetry~\cite{King:2013eh} to propose a new predictive neutrino mass model building scheme for the minimal seesaw model with two right-handed neutrinos
called the tri-direct CP approach~\cite{Ding:2018fyz,Ding:2018tuj}.

The basic idea of the tri-direct CP approach is that the Yukawa interactions associated with each of the two right-handed neutrinos preserve different residual flavour and CP symmetries, and the charged lepton sector also has a different residual
flavour symmetry. As a consequence, the flavour and generalized CP symmetry $G_{f}\rtimes H_{CP}$ is spontaneously broken down to $G_{l}$, $G_{\text{atm}}\rtimes H^{\text{atm}}_{CP}$ and $G_{\text{sol}}\rtimes H^{\text{sol}}_{CP}$ in the charged lepton, ``atmospheric'' and ``solar'' right-handed neutrino sectors, respectively~\cite{Ding:2018fyz}. Here $G_{l}$ is an abelian subgroup of $G_f$ and it allows the distinction of three generations of charged leptons as usual direct model. The residual subgroups $G_{\text{atm}}\rtimes H^{\text{atm}}_{CP}$ and $G_{\text{sol}}\rtimes H^{\text{sol}}_{CP}$ fix the alignments associated with each right-handed neutrino. We have performed a comprehensive analysis of lepton mixing patterns which can be obtained from the flavour group $S_4$ and CP symmetry in the tri-direct CP approach in a model independent fashion~\cite{Ding:2018tuj}. The model construction along the tri-direct CP approach was also illustrated~\cite{Ding:2018fyz,Ding:2018tuj}. In the minimal seesaw model, a phenomenologically viable pattern of lepton mixing and neutrino masses can also be obtained from the breaking of $A_5$ flavour symmetry into three different subgroups in the charged lepton, atmospheric neutrino and solar neutrino sectors~\cite{Ding:2017hdv}.

It is remarkable that the original Littlest Seesaw model for CSD(3) can be reproduced from the tri-direct CP approach~\cite{Ding:2018fyz,Ding:2018tuj}, if the $S_4$ flavour symmetry and CP symmetry are broken to the remnant symmetries $Z^T_3$, $Z^{U}_2\times H^{\text{atm}}_{CP}$ and $Z_2^{SU}\times H^{\text{sol}}_{CP}$ in the charged lepton sector, the atmospheric sector and the solar neutrino sector, respectively, corresponding to the
$\mathcal{N}_1$ case. In this case, one row of the neutrino Dirac mass matrix is proportional to $(0,-1,1)$ and the other row is proportional to $(1, 2-x, x)$, where $x$ is enforced to be a {\em real} parameter by the residual symmetry,
thereby overcoming the previous problem where it could be complex in general.  Then the light neutrino mass matrix is determined to be\footnote{Note that the seesaw mechanism results in a light effective Majorana mass matrix was defined in the convention ${\cal L_{\rm eff}}=- \frac{1}{2}\overline{\nu_{L}^c} m_{\nu}\nu_L  + \text{h.c.}$ Also note that here the second entries
of the vacuum alignments which enter the Dirac mass matrix
are multiplied by minus one as compared to the usual Littlest Seesaw convention.}~\cite{Ding:2018tuj}
\begin{equation}\label{eq:mnu}
 m_{\nu} =m_{a}\begin{pmatrix}
 0 &~ 0 ~& 0 \\
 0 &~ 1 ~& -1 \\
 0 &~ -1 ~& 1 \\
\end{pmatrix}+m_{s}e^{i\eta}
\begin{pmatrix}
 1 &~ 2-x &~ x \\
 2-x &~ (x-2)^2 &~ (2-x) x \\
 x &~ (2-x) x &~ x^2 \\
\end{pmatrix}\,,
\end{equation}
where an overall phase has been neglected, $m_{a}$, $m_{s}$, $\eta$ and $x$ are four real free parameters. In a concrete model, the parameters $x$ and $\eta$ could be fixed to certain values through the technique of vacuum alignment~\cite{Ding:2018fyz,Ding:2018tuj}. For example, CSD(3) corresponding to $x=3$
and $\eta=2\pi /3$, can be achieved within the $\mathcal{N}_1$ case.
Then all three mixing angles, two CP phases and three neutrino masses only depend on two real parameters $m_a$ and $m_s$ which can be determined by the mass squared differences $\Delta m^2_{21}\equiv m^2_2-m^2_1$ and $\Delta m^2_{31}\equiv m^2_3-m^2_1$ precisely measured in neutrino oscillation experiments. Then one can extract the predictions for all other mixing parameters. Obviously this kind of model is highly predictive.

In this paper, we shall focus on a particularly interesting example of the $\mathcal{N}_1$ case with
$x=-1/2$ and $\eta=-\pi/2$,
henceforth referred to as the new Littlest Seesaw,
which was one of the best fit points found in~\cite{Ding:2018tuj} where
the lepton mixing parameters and neutrino masses are predicted to lie in rather narrow regions, with an atmospheric angle in the second octant as preferred by the latest global fits. Motivated by the excellent agreement of this case with
experimental data, in this work we develop further this new Littlest Seesaw model in two different ways: we discuss leptogenesis and we also construct a concrete model to demonstrate how it could arise from a realistic theory.
We emphasise that the model involves a particularly simple and ``maximal'' phase $\eta=-\pi/2$ which is the unique source of CP violation for both neutrino oscillations and leptogenesis. It is noteworthy that the observed value of the baryon asymmetry $Y_{B}$ of our Universe will be obtained through flavoured thermal leptogenesis in both the SM and the Minimal Supersymmetric Standard Model (MSSM). We will propose an explicit
supersymmetric (SUSY)
model in the framework of minimal seesaw mechanism with 2RHN based on $S_{4}\rtimes H_{CP}$
and show that the mass hierarchies of the charged lepton and the light neutrino mass matrix in Eq.~\eqref{eq:mnu} with $x=-1/2$ and $\eta=-\pi/2$ may be naturally derived in such a model.

The rest of this paper is organized as follows. In section~\ref{sec:TDCP}, we revisit the $\mathcal{N}_1$ case of tri-direct CP models with the alignments $\langle\phi_{\text{atm}}\rangle\propto\left(0,1, -1\right)^T$, $\langle\phi_{\text{sol}}\rangle\propto\left(1, x, 2-x\right)^T$ which can be derived from the $S_4$ flavour symmetry in combination with CP symmetry, assuming the $\mathcal{N}_1$ residual symmetry. We show that the new Littlest Seesaw model, which corresponds to a benchmark point in the $\mathcal{N}_1$ case with $x=-1/2$ and $\eta=-\pi/2$, provides an excellent fit to the experimental data of lepton mixing angles and neutrino masses. We study the predictions of the new Littlest Seesaw model for leptogenesis in the section~\ref{sec:leptogenesis}, and show that the observed baryon asymmetry of the Universe can be produced for certain values of the lightest right-handed neutrino mass. In section~\ref{sec:model}, we construct a supersymmetric littlest tri-direct CP model based on the flavour symmetry $S_4\times Z_5\times Z_8$, the alignment parameter $x=-1/2$ and relative phase $\eta=-\pi/2$ are achieved. The predictions for the charged lepton flavour violation radiative decays $\l_i\rightarrow l_j\gamma$ are studied, and we show a UV completion of the model. In section~\ref{sec:Conclusion}, we summarize our main results and draw the conclusions. We present the group theory and the Clebsch-Gordan coefficients of the $S_4$ group in Appendix~\ref{sec:S4_group}.

\section{\label{sec:TDCP} The $\mathcal{N}_1$ case of tri-direct CP models revisited}

The tri-direct CP approach is based on the minimal seesaw model with 2RHN. We denote the two right-handed neutrinos as $N^c_{\text{atm}}$ (called ``atmospheric'') and $N^c_{\text{sol}}$ (called ``solar''). Then the most general Lagrangian of the minimal seesaw model can be written as
\begin{small}
\begin{equation}\label{eq:Lagrangian}
\mathcal{L}=-y_{l}L\phi_{l}E^{c}-y_{\text{atm}}L\phi_{\text{atm}}N^c_{\text{atm}}-y_{\text{sol}}L\phi_{\text{sol}}N^c_{\text{sol}}
-\frac{1}{2}x_{\text{atm}}\xi_{\text{atm}}N^c_{\text{atm}}N^c_{\text{atm}}-\frac{1}{2}x_{\text{sol}}\xi_{\text{sol}}N^{c}_{\text{sol}}N^c_{\text{sol}}
+\text{h.c.}\,,
\end{equation}
\end{small}
where two-component fermion notation for the fermion fields is adopted. The lepton doublets $L$ are assumed to transform as an irreducible triplet under $S_4$ ($L\sim \mathbf{3}$), $\phi_{\text{atm}}$ and $\phi_{\text{sol}}$ can be either Higgs fields
or combinations of the electroweak Higgs doublet together with flavons, and they are also $S_4$ triplets ($\phi_{\text{atm}}\sim \mathbf{3}$ and $\phi_{\text{sol}}\sim \mathbf{3^\prime}$). The two right-handed neutrinos are singlets of $S_4$ with $N^c_{\text{atm}}\sim\mathbf{1}$ and $N^c_{\text{sol}}\sim\mathbf{1^\prime}$, the two flavons $\xi_{\textrm{atm}}$ and $\xi_{\textrm{sol}}$ are invariant under $S_4$. The combination of flavons $\phi_{l}$ and the right-handed charged leptons $E^c\equiv(e^{c}, \mu^{c}, \tau^{c})^T$ must be embedded into the faithful three-dimensional representation $\mathbf{3}$ of $S_4$. Moreover, all coupling constants $y_{\text{atm}}$, $y_{\text{sol}}$, $x_{\text{atm}}$ and $x_{\text{sol}}$ are real because of the generalized CP symmetry.

We have performed an exhaustive analysis of all possible residual symmetries arising from $S_4\rtimes H_{CP}$ in tri-direct CP approach and the resulting predictions for neutrino masses and flavour mixing parameters in~\cite{Ding:2018tuj}. Many independent phenomenologically viable residual symmetry cases are found (eight cases for normal ordering and eighteen cases for inverted ordering). In the present work, we shall consider the breaking pattern in which the residual symmetries in the charged lepton, atmospheric neutrino and solar neutrino sectors are $Z^T_3$, $Z^{U}_2\times H^{\text{atm}}_{CP}$ and $Z_2^{SU}\times H^{\text{sol}}_{CP}$ respectively, the two residual CP symmetries are $H^{\text{atm}}_{CP}=\{1,U\}$ and $H^{\text{sol}}_{CP}=\{1,SU\}$. This is exactly the case $\mathcal{N}_1$ of Ref.~\cite{Ding:2018tuj}.
The residual symmetries in atmospheric neutrino and solar neutrino sectors require that the vacuum expectation values (VEVs) of the flavons  $\phi_{\textrm{atm}}$ and $\phi_{\textrm{sol}}$ should take the following form
\begin{equation}\label{eq:atm_sol_VEVs}
\langle\phi_{\text{atm}}\rangle=v_{\text{atm}}\left(0,1, -1\right)^T, \qquad \langle\phi_{\text{sol}}\rangle=v_{\text{sol}}\left(1, x, 2-x\right)^T\,,
\end{equation}
where the parameters $v_{\text{atm}}$, $v_{\text{sol}}$ and $x$ are real. Applying the well-known seesaw formula, the light neutrino mass matrix $m_{\nu}$ is really given by Eq.~\eqref{eq:mnu}.

In our working basis (see Appendix~\ref{sec:S4_group}),  requiring that the subgroup $Z^T_3$ is a symmetry of the charged neutrino mass matrix $m_l$ entails that $m^\dagger_lm_l$ is diagonal and thus does not contribute to the lepton mixing. The lepton mixing matrix is found to be of the following form~\cite{Ding:2018tuj}:
\begin{equation}\label{eq:PMNS}
U_{PMNS}=
\begin{pmatrix}
 \sqrt{\frac{2}{3}} &~ \frac{\cos \theta }{\sqrt{3}} &~ \frac{e^{i \psi } \sin \theta }{\sqrt{3}} \\
 -\frac{1}{\sqrt{6}} &~ \frac{\cos \theta }{\sqrt{3}}+\frac{e^{-i \psi } \sin \theta }{\sqrt{2}} &~ \frac{e^{i \psi } \sin \theta }{\sqrt{3}}-\frac{\cos \theta }{\sqrt{2}} \\
 -\frac{1}{\sqrt{6}} &~ \frac{\cos \theta }{\sqrt{3}}-\frac{e^{-i \psi } \sin \theta }{\sqrt{2}} &~ \frac{\cos \theta }{\sqrt{2}}+\frac{e^{i \psi } \sin \theta }{\sqrt{3}} \\
\end{pmatrix}P_{\nu}\,,
\end{equation}
where $P_{\nu}=\text{diag}(1, e^{i(\psi+\rho)/2}, e^{i(-\psi+\sigma)/2})$ is a diagonal phase matrix.
We see that the first column of the mixing matrix is in common with that of the tri-bimaximal mixing matrix, and the so-called TM1 mixing matrix is obtained. The neutrino mass spectrum is normal ordering, the lightest neutrino is massless ($m_1=0$) since only two right-handed neutrinos are involved. The other two non-zero light neutrino masses $m_2$ and $m_3$ are given by
\begin{equation}\label{eq:nu_masses}
m^2_2=\frac{m^2_a}{2}\left[9 r^2+w^2+12 r^2 (x-1)^2-\sqrt{B}\right], \quad
m^2_3=\frac{m^2_a}{2}\left[9 r^2+w^2+12 r^2 (x-1)^2+\sqrt{B}\right]\,,
\end{equation}
where
\begin{eqnarray}
\nonumber&&r=m_s/m_a, ~~ w=2 \sqrt{1+r^2 (x-1)^4+2 r (x-1)^2 \cos \eta },\\
\nonumber&&B=\left(9 r^2-w^2\right)^2+24  r^2 (x-1)^2A\,,\\
\label{eq:theta}&&A=9 r^2+w^2+6 r w \cos (\eta -\phi_{w}),~~  \phi_{w}=\text{arg}\left(1+r (x-1)^2 e^{ i \eta }\right) \,.
\end{eqnarray}
The expressions for the angles and phases $\theta$, $\psi$, $\rho$ and $\sigma$ in Eq.~\eqref{eq:PMNS} are:
\begin{eqnarray}
\nonumber && \cos2\theta=\frac{w^2-9 r^2}{\sqrt{B}}, \qquad \sin2\theta=\frac{2 \sqrt{6A}  r (x-1)}{\sqrt{B}}, \qquad \sin\psi=-\frac{w \sin (\eta -\phi_{w})}{\sqrt{A}}\,, \\
\label{eq:prs_prs}&&\cos\psi=\frac{3 r+w \cos (\eta -\phi_{w})}{\sqrt{A}}, \qquad \sin(\rho-\sigma)=\frac{3r wm^2_a \sqrt{B}  \sin (\eta -\phi_{w})}{m_2m_3A }\,.
\end{eqnarray}
From the lepton mixing matrix in Eq.~\eqref{eq:PMNS}, one can straightforwardly extract the following results for the lepton mixing angles and CP invariants,
\begin{eqnarray}
\nonumber &&\sin^2\theta_{13}=\frac{\sin ^2\theta }{3}=\frac{1}{6}\left(1-\frac{w^2-9 r^2}{\sqrt{B}}\right)\,, \qquad \sin^2\theta_{12}=\frac{2 \cos ^2\theta }{5+\cos 2 \theta }=\frac{1}{3}\left(1-2\tan^2\theta_{13}\right)\,,\\
\nonumber&& \sin^2\theta_{23}=\frac{1}{2}-\frac{ \sqrt{6} \sin 2 \theta  \cos \psi  }{5+\cos 2 \theta}=\frac{1}{2}-\frac{12  r (x-1)[3 r+w \cos (\eta -\phi_{w})]  }{5\sqrt{B}+w^2-9 r^2}\,, \\
 \nonumber && J_{CP}=\frac{\sin 2\theta   \sin \psi }{6 \sqrt{6}}=-\frac{w  r (x-1)\sin (\eta -\phi_{w})}{3 \sqrt{B}}\,, \\
\label{eq:mix_pars}&& I_{1}=\frac{1}{36} \sin ^22 \theta  \sin (\rho -\sigma )=\frac{2 r^3 w (x-1)^2 \sin (\eta -\phi_{w})}{m_2m_3\sqrt{B} }\,,
\end{eqnarray}
\begin{table}[t!]
\renewcommand{\tabcolsep}{0.5mm}
\renewcommand{\arraystretch}{1.3}
\footnotesize
\centering
\begin{tabular}{|c| c| c|c| c | c| c| c| c| c| c |c | c| c| c|}  \hline \hline
$\langle\phi_{\text{sol}}\rangle/v_{\phi_{s}}$ & $x$   & $\eta$ & $m_{a}$(meV) & $r$  	 & $\chi^2_{\text{min}}$ &  $\sin^2\theta_{13}$  &$\sin^2\theta_{12}$  & $\sin^2\theta_{23}$  & $\delta_{CP}/\pi$ &  $\varphi/\pi$  & $m_2(\text{meV})$ & $m_3(\text{meV})$ & $m_{ee}(\text{meV})$\\   \hline
$\left(1,3,-1\right)^{T}$ & $3$ & $\pm\frac{2 \pi }{3}$ & $26 .850$ & $0 .0997$ & $24 .861$ & $0 .0221$ & $0 .318$ & $0 .488$ & $\mp0.516$ & $\mp0.403$ & $8 .579$ & $50 .272$ & $2 .677$ \\ \hline
$\left(1,-1,3\right)^{T}$ & $-1$ & $\pm\frac{2 \pi }{3}$ & $26 .796$ & $0 .101$ & $13 .744$ & $0 .0225$ & $0 .318$ & $0 .513$ & \
$\pm0 .482$ & $\mp0.401$ & $8 .632$ & $50 .210$ & $2 .696$ \\ \hline
$\left(1,4,-2\right)^{T}$ &$4$ & $\pm\frac{4 \pi }{5}$ & $35 .249$ & $0 .0564$ & $14 .358$ & $0 .0241$ & $0 .317$ & $0 .575$ & $\mp0.398$ & $\mp0.474$ & $8 .315$ & $50 .610$ & $1 .990$ \\ \hline
\multirow{2}{*}{$\left(1,\frac{7}{2},-\frac{3}{2}\right)^{T}$} & \multirow{2}{*}{$\frac{7}{2}$} & $\pm\frac{3 \pi }{4}$ & $31 .123$ & $0 .0734$ & $7 .823$ & $0 .0231$ & $0 .318$ & $0 .541$ & $\mp0.444$ & $\mp0.447$ & $8 .459$ & $50 .429$ & $2 .284$\\ \cline{3-14}
&& $\pm\frac{4 \pi }{5}$ & $33 .016$ & $0 .0673$ & $9 .143$ & $0 .0209$ & $0 .319$ & $0 .589$ & $\mp0.366$ & $\mp0.544$ & $8 .802$ & $50 .014$ & $2 .222$ \\ \hline
$\left(1,\frac{10}{3},-\frac{4}{3}\right)^{T}$ & $\frac{10}{3}$ & $\pm\frac{3 \pi }{4}$ & $30 .572$ & $0 .0777$ & $5 .183$ & $0 .0218$ & $0 .318$ & $0 .548$ & $\mp0.432$ & $\mp0.474$ & $8 .685$ & $50 .150$ & $2 .374$ \\ \hline
\rowcolor{LightCyan}
$\left(1,-\frac{1}{2},\frac{5}{2}\right)^{T}$ & $-\frac{1}{2}$ &  $\pm\frac{\pi }{2}$ &  $22 .366$ & $0 .145$ &  $2 .487$ & $0 .0220$ & $0 .318$ & $0 .599$ & $\pm0 .354$ & $\mp0.317$ & $8 .670$ &  $50 .167$ & $3 .241$ \\ \hline
$\left(1,-\frac{2}{3},\frac{8}{3}\right)^{T}$ & $-\frac{2}{3}$ & $\pm\frac{3 \pi }{5}$ & $24 .571$ & $0 .122$ & $14 .594$ & $0 .0212$ & $0 .319$ & $0 .545$ & $\pm0 .435$ & $\mp0.383$ & $8 .889$ & $49 .911$ & $3 .009$ \\ \hline
$\left(1,-\frac{3}{4},\frac{11}{4}\right)^{T}$ & $-\frac{3}{4}$ & $\pm\frac{3 \pi }{5}$ & $24 .579$ & $0 .120$ & $3 .600$ & $0 .0222$ & $0 .318$ & $0 .551$ & $\pm0 .429$ & $\mp0.367$ & $8 .670$ & $50 .167$ & $2 .949$  \\ \hline
$\left(1,-\frac{3}{5},\frac{13}{5}\right)^{T}$ & $-\frac{3}{5}$ & $\pm\frac{\pi }{2}$ & $22 .220$ & $0 .142$ & $11 .666$ & $0 .0232$ & $0 .318$ & $0 .606$ & $\pm0 .347$ & $\mp0.297$ & $8 .309$ & $50 .618$ & $3 .155$  \\ \hline
$\left(1,-\frac{4}{5},\frac{14}{5}\right)^{T}$ & $-\frac{4}{5}$ & $\pm\frac{3 \pi }{5}$ & $24 .585$ & $0 .118$ & $3 .249$ & $0 .0228$ & $0 .318$ & $0 .554$ & $\pm0 .425$ & $\mp0.357$ & $8 .534$ & $50 .333$ & $2 .912$ \\ \hline
$\left(1,-\frac{5}{6},\frac{17}{6}\right)^{T}$ & $-\frac{5}{6}$ & $\pm\frac{3 \pi }{5}$ & $24 .590$ & $0 .117$ & $5 .588$ & $0 .0231$ & $0 .318$ & $0 .556$ & $\pm0 .422$ & $\mp0.350$ & $8 .443$ & $50 .451$ & $2 .887$ \\ \hline \hline

\end{tabular}
\caption{\label{tab:bf}Some benchmark values of the parameters $x$ and $\eta$ and the corresponding predictions for the lepton mixing angles, CP violation phases, neutrino masses and the effective Majorana mass $m_{ee}$.
These results are benchmark examples in the $\mathcal{N}_1$ class of tri-direct CP models~\cite{Ding:2018tuj}. Notice that the lightest neutrino mass is vanishing $m_1=0$. }
\end{table}
where $J_{CP}$ is the Jarlskog invariant~\cite{Jarlskog:1985ht} and $I_1$ is the Majorana invariant~\cite{Branco:1986gr} related to the Majorana phase $\varphi$. We find that all mixing parameters and mass ratio $m_2/m_3$ depend on the three input parameters $x$, $\eta$ and $r=m_s/m_a$. However, the neutrino absolute masses $m_2$ and $m_3$ depend on all the four input parameters $x$, $\eta$, $m_a$ and $r$. We find that the agreement with data is optimised by choosing
\begin{equation}
m_{a}=23.133\,\text{meV}, \quad r=0.135, \quad \eta=-0.542\pi,\quad  x=-0.615\,,
\end{equation}
which give rise to the following values of observables
\begin{eqnarray}
\nonumber &&\sin^2\theta_{13}=0.02241, \quad \sin^2\theta_{12}=0.318, \quad \sin^2\theta_{23}=0.582, \quad \delta_{CP}=-0.382\pi,\quad \varphi=0.333\pi\,, \\
&&m_1=0\,\text{meV}, \qquad m_2=8.597\,\text{meV}, \qquad m_3=50.249\,\text{meV}, \qquad m_{ee}=3.112\,\text{meV}\,,
\end{eqnarray}
where $m_{ee}$ refers to the effective Majorana mass in neutrinoless double beta decay, and $\varphi$ is the Majorana phase. These predictions for lepton mixing angles agree with the experimental data quite well, and the global minimum of the $\chi^2$ function is $\chi^2_{\text{min}}=0.384$. Note that the $\chi^2$ function includes the contributions of three mixing angles and two squared mass differences as usual. Because the indication of a preferred value of the Dirac phase $\delta_{CP}$ from global data analyses is rather weak~\cite{Esteban:2018azc}, we do not include any information on $\delta_{CP}$ in the $\chi^2$ function.
We emphasise that the values of the parameter $x$, $\eta$, $r$ and $m_a$ are not fixed by the residual symmetry,
and can only be fixed by explicit model construction. This task is easier for
the simpler values of $x$ and $\eta$ where the solar vacuum alignment $\langle\phi_{\text{sol}}\rangle$ is easier to achieve, therefore we are interested in the simplest values of these parameters.

We report the results of $\chi^2$ analysis for some representative values of $x$ and $\eta$ in table~\ref{tab:bf}. Once the values of $x$ and $\eta$ are fixed, all the mixing parameters and neutrino masses only depend on the input parameters $m_a$ and $r$ whose values can be determined by the mass squared differences $\Delta m^2_{21}$ and $\Delta m^2_{31}$. Then the three lepton mixing angles, two CP violation phases and the absolute neutrino mass scale are uniquely predicted by the theory. We notice that the effective Majorana mass $m_{ee}$ lies in the range of 1 to 4 meV, consequently it is impossible to be measured in foreseeable future.

\begin{figure}[t!]
\centering
\begin{tabular}{c}
\includegraphics[width=0.6\linewidth]{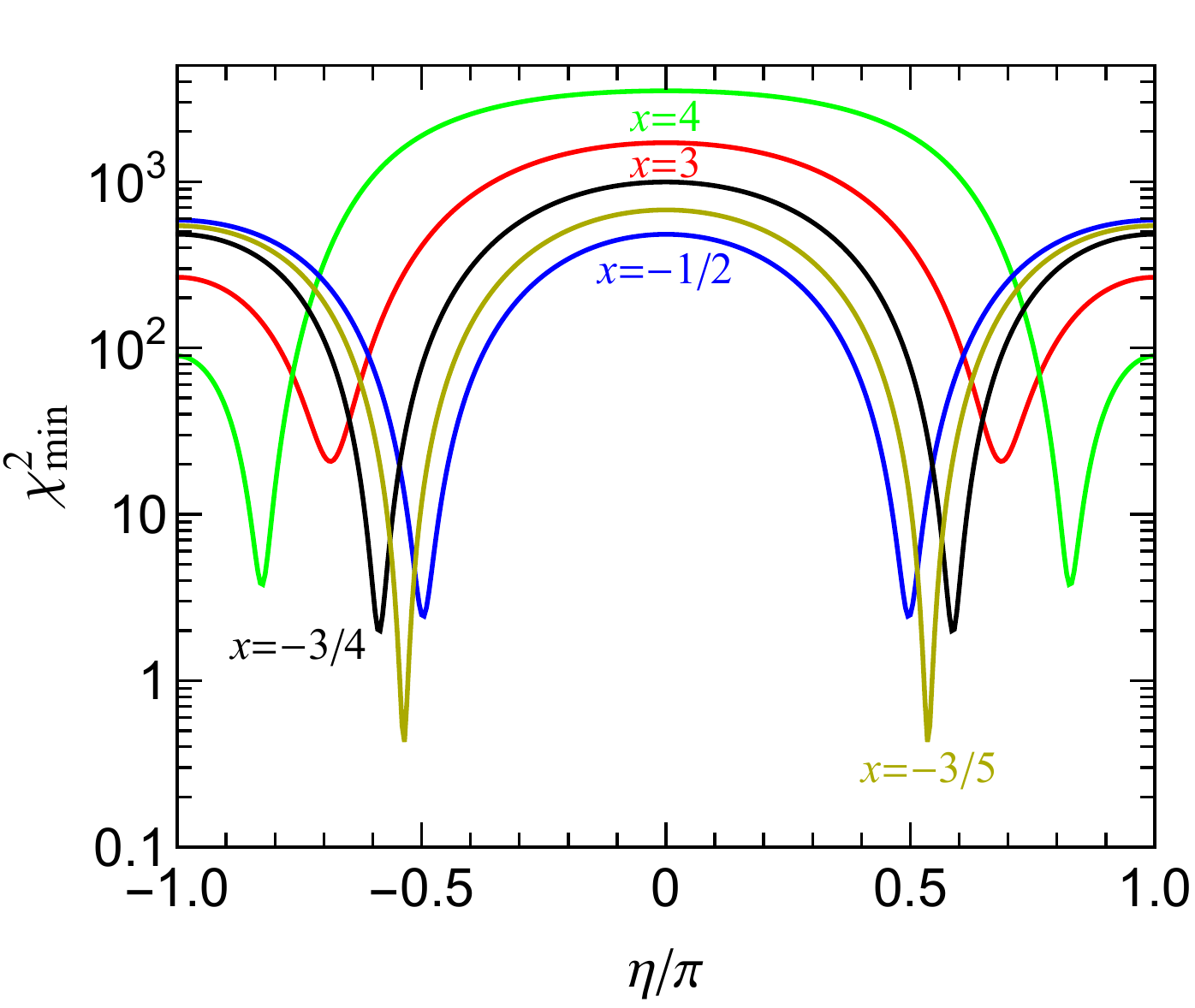}
\end{tabular}
\caption{\label{fig:chi2_vs_eta} Variation of $\chi^2$ with respect to the phase $\eta$ for the typical values of $x=3, 4, -1/2,-3/4, -3/5$, for the
$\mathcal{N}_1$ case of tri-direct CP models.}
\end{figure}

\begin{figure}[t!]
\centering
\begin{tabular}{c}
\includegraphics[width=0.99\linewidth]{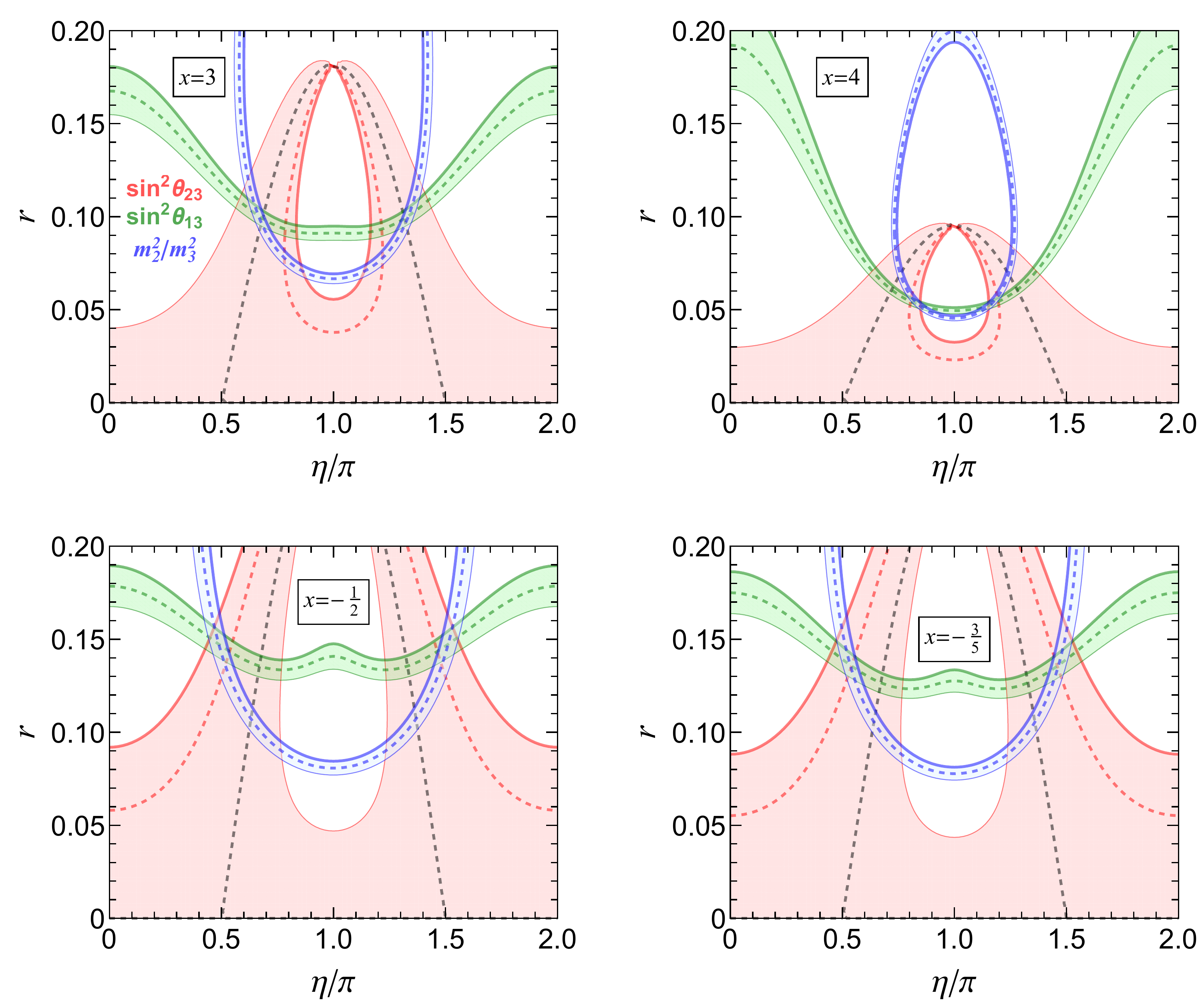}
\end{tabular}
\caption{\label{fig:contour_mix_par} Contour plots of $\sin^2\theta_{13}$, $\sin^2\theta_{23}$ and $m_2/m_3$ in the $\eta/\pi-r$ plane for $x=3$, $4$, $-1/2$ and $-3/5$, for the
$\mathcal{N}_1$ case of tri-direct CP models. The red, green and blue areas denote the $3\sigma$ contour regions of $\sin^2\theta_{23}$, $\sin^2\theta_{13}$ and the mass ratio $m^2_2/m^2_3$ respectively.
The dashed lines denote the best fit values from NuFIT 4.0.}
\end{figure}

The original Littlest Seesaw model~\cite{King:2015dvf,King:2016yvg,Ballett:2016yod,King:2018fqh} corresponds to the cases of $(x, \eta)=(3, 2\pi/3)$, $(-1, -2\pi/3)$, and the CSD(4) model~\cite{King:2013xba,King:2013hoa} can be exactly reproduced for $(x, \eta)=(4, 4\pi/5)$. From table~\ref{tab:bf}, we see that the values $(x, \eta)=(-1/2, \pm\pi/2)$, $(-3/4, \pm3\pi/5)$ and $(-4/5, \pm3\pi/5)$ can give rise to a smaller $\chi^2_{\text{min}}$ than the original Littlest Seesaw model and CSD(4) model~\cite{King:2015dvf,King:2016yvg,Ballett:2016yod,King:2018fqh,King:2013xba,King:2013hoa}. We have shown $\chi^2_{\text{min}}$ as a function of $\eta$ for $x=3, 4, -1/2,-3/4, -3/5$ in figure~\ref{fig:chi2_vs_eta}. Moreover, we plot the contour regions for the $3\sigma$ intervals of mixing angles $\theta_{13}$ and $\theta_{23}$ and mass ratio
$m_2/m_3$ in the plane $r$ versus $\eta/\pi$ in figure~\ref{fig:contour_mix_par}.  The result for $\theta_{12}$ is not displayed here, because it is related to the reactor angle $\theta_{13}$ by the TM1 mixing sum rule $\cos^2\theta_{12}\cos^2\theta_{13}=2/3$ which leads to
$0.316\leq\sin^2\theta_{12}\leq0.319$ for the $3\sigma$ allowed range of $\theta_{13}$~\cite{Esteban:2018azc}.

From figures~\ref{fig:chi2_vs_eta} and~\ref{fig:contour_mix_par}, we notice that the values of $\chi^2_{\text{min}}$ is quite sensitive to the phase $\eta$ and predictions for the mixing angles and neutrino masses can agree very well with the experimental data for certain choices of $\eta$. Henceforth we shall focus on the new Littlest Seesaw model defined by
the simple values $x=-1/2$ and $\eta=\pm\pi/2$ which give a phenomenologically successful and predictive description of lepton mixing parameters and neutrino masses, as highlighted with cyan background in table~\ref{tab:bf}. Moreover, the corresponding vacuum alignment $\langle\phi_{\text{sol}}\rangle\propto \left(1,-1/2,5/2\right)$ and the phase $\eta=\pm\pi/2$ should be easy to realize in a concrete model. This new Littlest Seesaw model and the original Littlest Seesaw model differ in their predictions for $\theta_{23}$ and $\delta_{CP}$. The atmospheric mixing angle $\theta_{23}$ deviates from maximal mixing in the new Littlest Seesaw model while it is close to $45^{\circ}$ in the original Littlest Seesaw. Since deviation of $\theta_{23}$ from maximal mixing is preferred by the present data~\cite{Esteban:2018azc}, the new littlest tri-direct CP model provides a better fit to the data of $\theta_{23}$ than the original Littlest Seesaw.

\subsection{\label{sec:spec_mix} The new Littlest Seesaw: $x=-1/2$, $\eta=-\pi/2$ }
Before getting into too many technicalities of model construction, we analyze the predictions for lepton mixing parameters and neutrino masses for $x=-1/2, \eta=-\pi/2$. In this case, the light neutrino mass matrix in Eq.~\eqref{eq:mnu} becomes
\begin{equation}
\label{eq:mnu_model}m_{\nu}=m_a\begin{pmatrix}
 0 &~ 0 ~& 0 \\
 0 &~ 1 ~& -1 \\
 0 &~ -1 ~& 1 \\
\end{pmatrix}-\frac{im_s}{4}\begin{pmatrix}
 4 &~ 10 &~ -2 \\
 10 &~ 25 &~ -5 \\
 -2 &~ -5 &~ 1 \\
\end{pmatrix}\,.
\end{equation}
We note that all lepton mixing parameters and mass ratio $m_2/m_3$ are determined by only a single parameter $r=m_s/m_a$. The expressions for the three lepton mixing angles and the CP invariants are given by
\begin{eqnarray}
\nonumber &&\sin^2\theta_{13}=\frac{1}{6} \left(1-\frac{45 r^2+16}{C_r}\right)\,, \qquad \sin^2\theta_{12}=1-\frac{4 C_{r}}{5 C_{r}+45 r^2+16}\,, \\
 &&\sin^2\theta_{23}=\frac{1}{2}+\frac{540 r^2}{5 C_{r}+45 r^2+16}\,,\qquad J_{CP}=-\frac{4r}{ C_{r}}, \qquad I_{1}=-\frac{6r^2}{ C_{r}}\,,
\end{eqnarray}
with
\begin{equation}\label{eq:Cr}
C_{r}=4\sqrt{B}\left.\right|_{x=-1/2,\eta=-\pi/2}=\sqrt{\left(225 r^2+16\right)^2-2304 r^2}\,.
\end{equation}
Notice that $\theta_{23}$ is predicted to lie in the second octant, it is preferred by the present neutrino oscillation data~\cite{Esteban:2018azc}. As both $\theta_{13}$ and $\theta_{23}$ depend on a single parameter $r$, a sum rule between them can be obtained\footnote{The sum rule for $\theta_{12}$ is $\cos^2\theta_{12}\cos^2\theta_{13}=2/3$ which holds true for all TM1 models.}
\begin{equation}
\sin^2\theta_{23}=\frac{1+4 \sin ^2\theta_{13}+\sqrt{1+28 \sin ^2\theta_{13}  (1-3 \sin ^2\theta_{13} )}}{4 \cos ^2\theta_{13}}\,.
\end{equation}
The two non-zero neutrino masses can be read off from Eq.~\eqref{eq:nu_masses} as,
\begin{equation}
m^2_2=\frac{1}{8}m^2_a  \left(16+225 r^2- C_{r}\right)\,, \qquad
m^2_3=\frac{1}{8}m^2_a  \left(16+225 r^2+ C_{r}\right) \,.
\end{equation}
It is easy to see that the mass ratio $m_2/m_3$ only depends on the parameter $r$. Consequently we can express the mass ratio $m^2_2/m^2_3$ in terms of $\theta_{13}$ as
\begin{equation}
\frac{m^2_2}{m^2_3}=\frac{10 \sin ^2\theta_{13}  (3 \sin ^2\theta_{13} -1)+\sqrt{1+28 \sin ^2\theta_{13}  (1-3 \sin ^2\theta_{13} )}-1}{2 \sin ^2\theta_{13}  (15 \sin ^2\theta_{13} -8)+2}\,.
\end{equation}
\begin{figure}[t!]
\centering
\begin{tabular}{cc}
\includegraphics[width=0.48\linewidth]{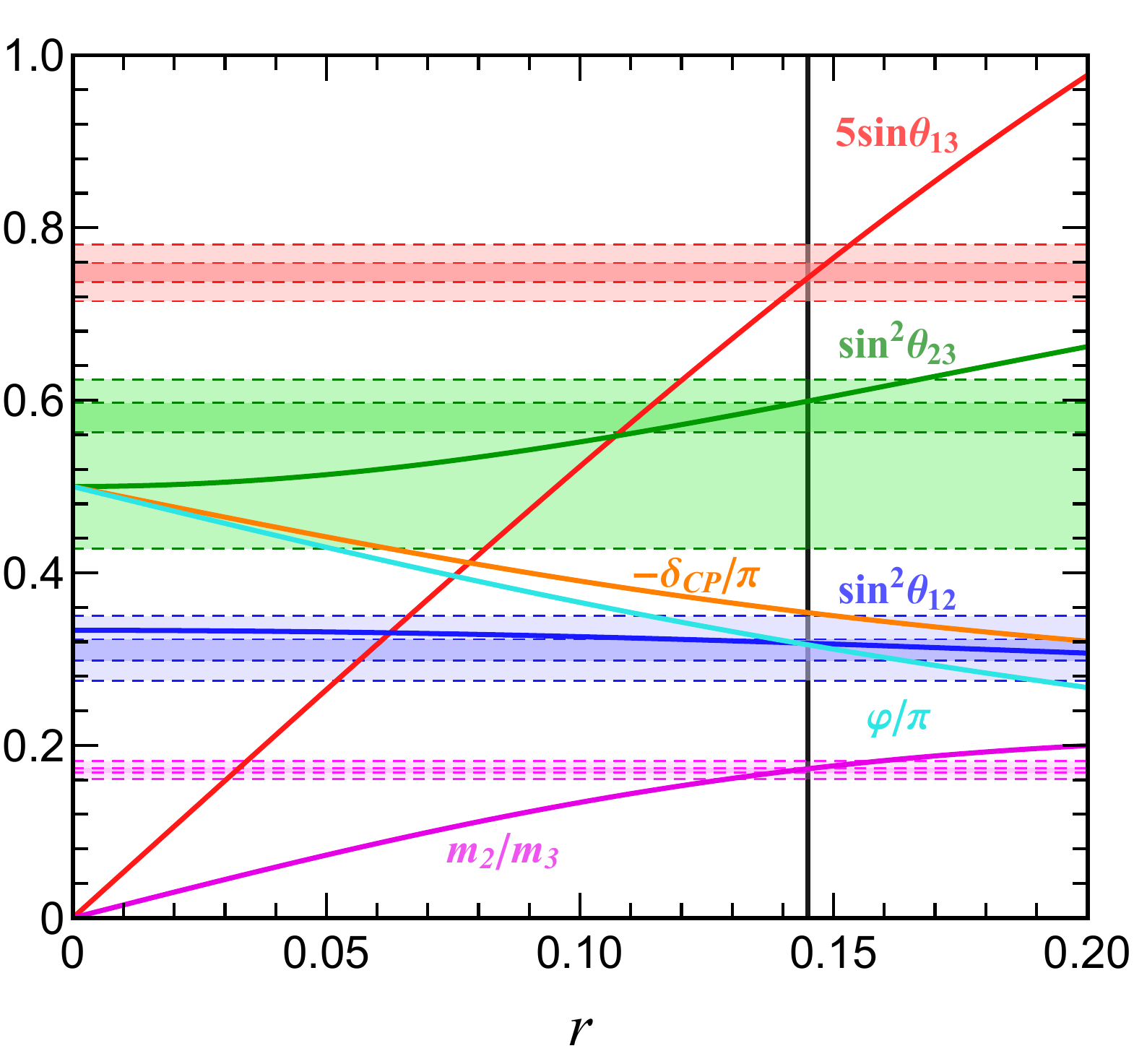} &
\includegraphics[width=0.48\linewidth]{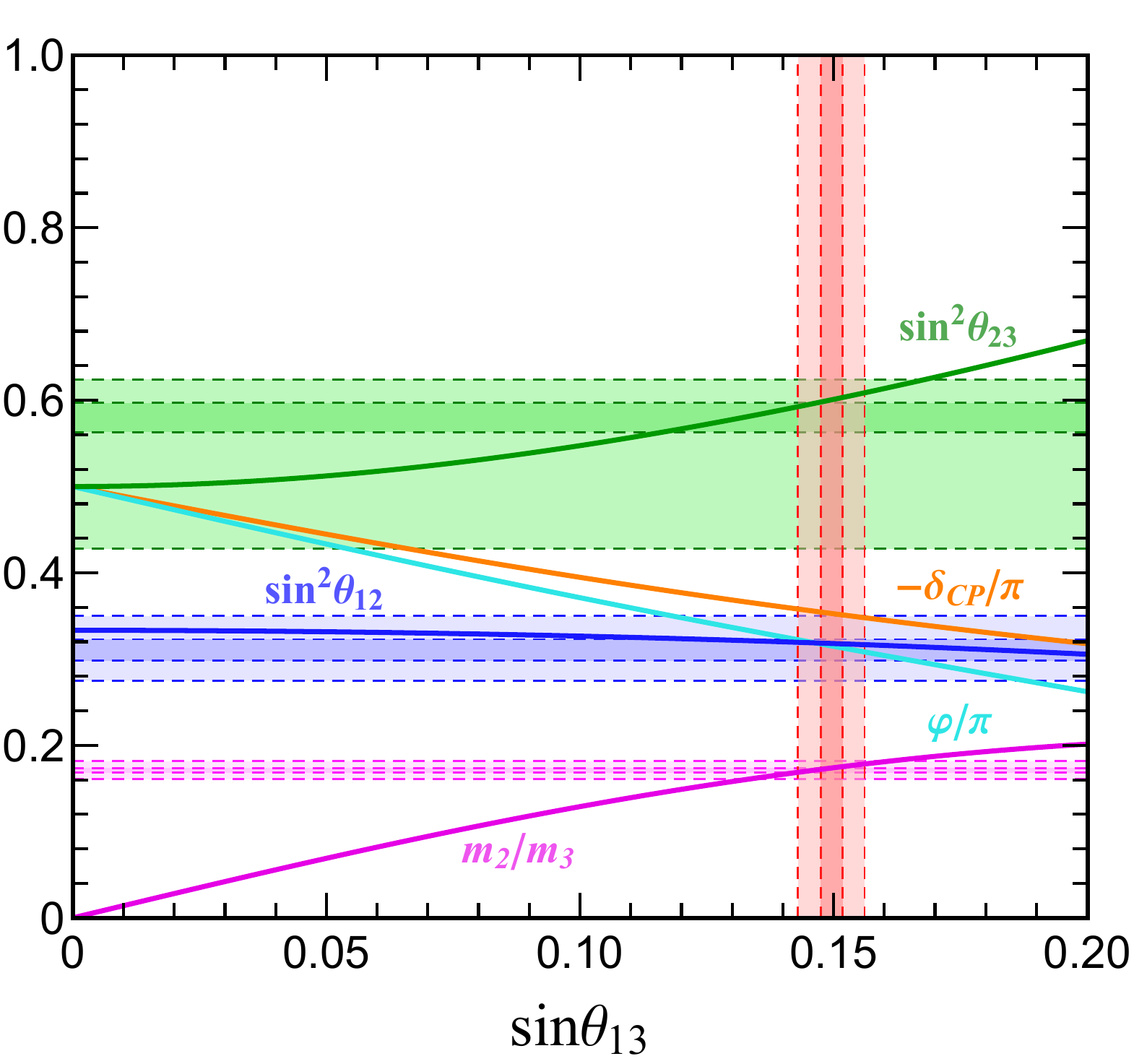}
\end{tabular}
\caption{\label{fig:Mix_par} The predictions of the new Littlest Seesaw model with $x=-1/2$, $\eta=-\pi/2$ for the mixing parameters and mass ratio $m_2/m_3$. The shaded regions represent the $1\sigma$ and $3\sigma$ ranges of each mixing parameter and mass ratio~\cite{Esteban:2018azc}. On the left panel, the values of mixing parameters and mass ratio are predicted with respect to $r$ and the black vertical line denotes the best fit value $r_{\text{bf}}=0.145$. On the right panel, we show the predictions for mixing parameters and mass ratio as functions of $\sin\theta_{13}$.
}
\end{figure}
We plot the dependence of all lepton mixing parameters and mass ratio $m_2/m_3$ on the parameter $r$ in figure~\ref{fig:Mix_par}. Eliminating the input parameter $r$, we can relate all above physical observables to the reactor mixing angle $\theta_{13}$. We see from  figure~\ref{fig:Mix_par} that the three lepton mixing angles and neutrino mass ratio are within their $1\sigma$ ranges at the best fit point $r=0.145$. The best fitting values of Dirac CP phase and Majorana CP phase are $\delta_{CP}\simeq-0.354\pi$ and $\varphi\simeq0.316\pi$, respectively. We numerically scan over the parameter space of $m_a$ and $r$, and find the viable range of $r$ is $r\in[0.139,0.153]$ to be compatible with the present neutrino oscillation data at $3\sigma$ level~\cite{Esteban:2018azc}. Furthermore, we find the neutrino masses and mixing parameters are predicted to lie in the following rather narrow regions,
\begin{eqnarray}
\nonumber &&0.3167\leq \sin^2\theta_{12}\leq0.3194, \qquad 0.02044\leq \sin^2\theta_{13}\leq0.02437, \qquad 0.593\leq \sin^2\theta_{23}\leq0.609, \\
\nonumber &&-0.358\leq\delta_{CP}/\pi\leq-0.348, \qquad  0.308\leq\varphi/\pi\leq0.322, \qquad  3.084\,\text{meV}\leq m_{ee}\leq3.388\,\text{meV}, \\
\label{eq:model_prediction} &&8.319\,\text{meV}\leq m_2\leq8.950\,\text{meV}, \qquad 49.305\,\text{meV}\leq m_3\leq51.206\,\text{meV}\,.
\end{eqnarray}
Therefore this new Littlest Seesaw model is very predictive and it should be easily excluded by precise measurement of $\theta_{12}$, $\theta_{23}$ and $\delta_{CP}$ in forthcoming neutrino facilities.

\subsection{The new Littlest Seesaw as a limiting case of three right-handed neutrinos}

We shall extend the idea of Littlest seesaw to the 3RHN model in the following. We denote the 3RHN as $N^c_{\text{atm}}$, $N^c_{\text{sol}}$ and $N^c_{\text{dec}}$. Then for the seesaw Lagrangian in Eq.~\eqref{eq:Lagrangian}, the two additional terms related to the third right-handed neutrino $N^c_{\text{dec}}$ can be written as
\begin{equation}\label{eq:Lagrangian_extens}
\nonumber\Delta\mathcal{L}=-y_{\text{dec}}L\phi_{\text{dec}}N^c_{\text{dec}}
-\frac{1}{2}x_{\text{dec}}\xi_{\text{dec}}N^{c}_{\text{dec}}N^c_{\text{dec}}
+\text{h.c.}\,.
\end{equation}
Here the flavon $\phi_{\text{dec}}$ is assigned to transform as $S_4$ triplet $\mathbf{3}$, both $\xi_{\text{dec}}$ and $N^{c}_{\text{dec}}$ are invariant under the actions of $S_4$. As an example, we consider the case that the residual symmetry in the decoupled neutrino sector is $Z^{T^2ST}_{2}\times H^{\text{dec}}_{CP}$ with $H^{\text{dec}}_{CP}=\{SU,TST^2U\}$. Then the most general VEV of $\phi_{\text{dec}}$ which preserves the above residual symmetry is
\begin{equation}
 \langle\phi_{\text{dec}}\rangle \propto\left(1, \omega, \omega^2\right)^T\,.
\end{equation}
Then the neutrino mass matrix in Eq.~\eqref{eq:mnu_model} becomes
\begin{equation}
 m_{\nu}=m_{a}\begin{pmatrix}
 0 &~ 0 &~ 0 \\
 0 &~ 1 &~ -1 \\
 0 &~ -1 &~ 1
\end{pmatrix}-\frac{i}{4}m_{a}r
\begin{pmatrix}
 4 &~ 10 &~ -2 \\
 10 &~ 25 &~ -5 \\
 -2 &~ -5 &~ 1
\end{pmatrix}+m_{a}r^\prime e^{i\eta^\prime}
\begin{pmatrix}
1 &~  \omega ^2 &~  \omega  \\
  \omega ^2 &~  \omega  &~ 1 \\
  \omega  &~ 1 &~ \omega ^2
\end{pmatrix}\,,
\end{equation}
where $m_{a}$, $r$, $r^\prime$, $\eta$ and $\eta^\prime$ are real. The first two terms coincide with those of the new Littlest Seesaw in Eq.~\eqref{eq:mnu_model}, and the last term arises from the third decoupled right-handed neutrinos. An particularly interesting example is the case of $\eta^\prime=0$, it predicts the best fit values of the mixing parameters as follows
\begin{eqnarray}
\nonumber &&\hskip-0.15in  m_{a}=22.663\,\text{meV}, \quad  r=0.141, \quad r^\prime=0.00834, \quad\eta=-\pi/2,\quad \eta^\prime=0, \quad \chi^2_{\text{min}}=1.157\,,\\
\nonumber && \hskip-0.15in  \sin^2\theta_{13}=0.0224,\quad  \sin^2\theta_{12}=0.318, \quad \sin^2\theta_{23}=0.595, \quad \delta_{CP}=-0.363\pi, \quad \alpha_{21}=0.394\pi\,, \\
\label{eq:3RHN_bf}&& \hskip-0.15in  \alpha_{31}=0.0716\pi,\quad   m_1=0.285\,\text{meV}, \quad m_2=8.577\,\text{meV},  \quad m_3=50.283, \quad m_{ee}=3.197\,\text{meV}\,.
\end{eqnarray}
We see $r'\ll r\ll 1$ such that the condition of constrained sequence dominance is well satisfied. Therefore our new Littlest Seesaw with 2RHN  can be regarded as a decoupling limit of the 3RHN model in the case of $M_{\rm dec}\gg M_{\rm atm}, M_{\rm sol}$.
Comparing the best fit values of 3RHN model in Eq.~\eqref{eq:3RHN_bf} with  those of 2RHN model with $x=-1/2$ and $\eta=-\pi/2$ in table~\ref{tab:bf}, we find that the 2RHN model is a good approximation of the 3RHN model.

\section{\label{sec:leptogenesis}Predictions for leptogenesis in the new Littlest Seesaw model}

It is well-known fact that there is a predominance of matter over antimatter present in the observable Universe. The value of baryon asymmetry of the Universe normalised to the entropy density is~\cite{Aghanim:2018eyx},
\begin{equation} \label{eq:YB obs}
Y_{B}=(0.870300\pm 0.011288)\times10^{-10} \quad (95\% \text{CL})\,.
\end{equation}
Apart from elegantly explaining the tiny neutrino masses, the seesaw mechanism provides a simple and attractive mechanism for understanding the matter-antimatter asymmetry of the Universe via leptogenesis~\cite{Fukugita:1986hr}. The out-of-equilibrium decays of right-handed neutrinos in the early Universe generates a lepton asymmetry because of the CP violating Yukawa couplings. The lepton asymmetry is subsequently converted into a baryon asymmetry via sphaleron processes in the SM.

In our concerned model, the phase $\eta$ is the unique source of CP violation, and it controls CP violation in both neutrino oscillations and leptogenesis. Therefore the measurable CP violation in future neutrino oscillation experiments are closely related to the baryon asymmetry of the Universe. In the present work, we shall focus on the simplest version of the leptogenesis in which the lepton asymmetry is dominantly generated by the interactions and decay of the lightest right-handed neutrino. The phase $\eta$ is fixed to $\eta=-\pi/2$ in the new Littlest Seesaw model, and it yields a Dirac CP violation phase $\delta_{CP}\simeq1.646\pi$. In this section, we shall study the prediction for leptogenesis within the framework of SM and MSSM.
The condition of successful baryogenesis will allow us to determine the mass of the lightest right-handed neutrino in the new Littlest Seesaw model.

\subsection{\label{sec:Lep_SM} Leptogenesis for the new Littlest Seesaw model in the SM  }

In the SM, the final baryon asymmetry is given by~\cite{Nardi:2006fx}
\begin{equation}
  \label{eq:YB SM}
  Y_{B}=\frac{12}{37} \sum_{\alpha} Y_{\Delta_\alpha}\,,
\end{equation}
where the asymmetries $Y_{\Delta_\alpha}$ ($\alpha=e,\mu,\tau$) are defined as $Y_{\Delta_\alpha}\equiv Y_B/3 - Y_{L_\alpha}$ and they are conserved by the sphaleron processes~\cite{Abada:2006ea}. $Y_{L_\alpha}$ refers to the lepton number densities of the flavour $\alpha$. Note that $Y_{B}$, $Y_{\Delta_\alpha}$ and $Y_{L_\alpha}$ is normalised to the entropy density.

\begin{figure}[t!]
\centering
\begin{tabular}{c}
\includegraphics[width=0.60\linewidth]{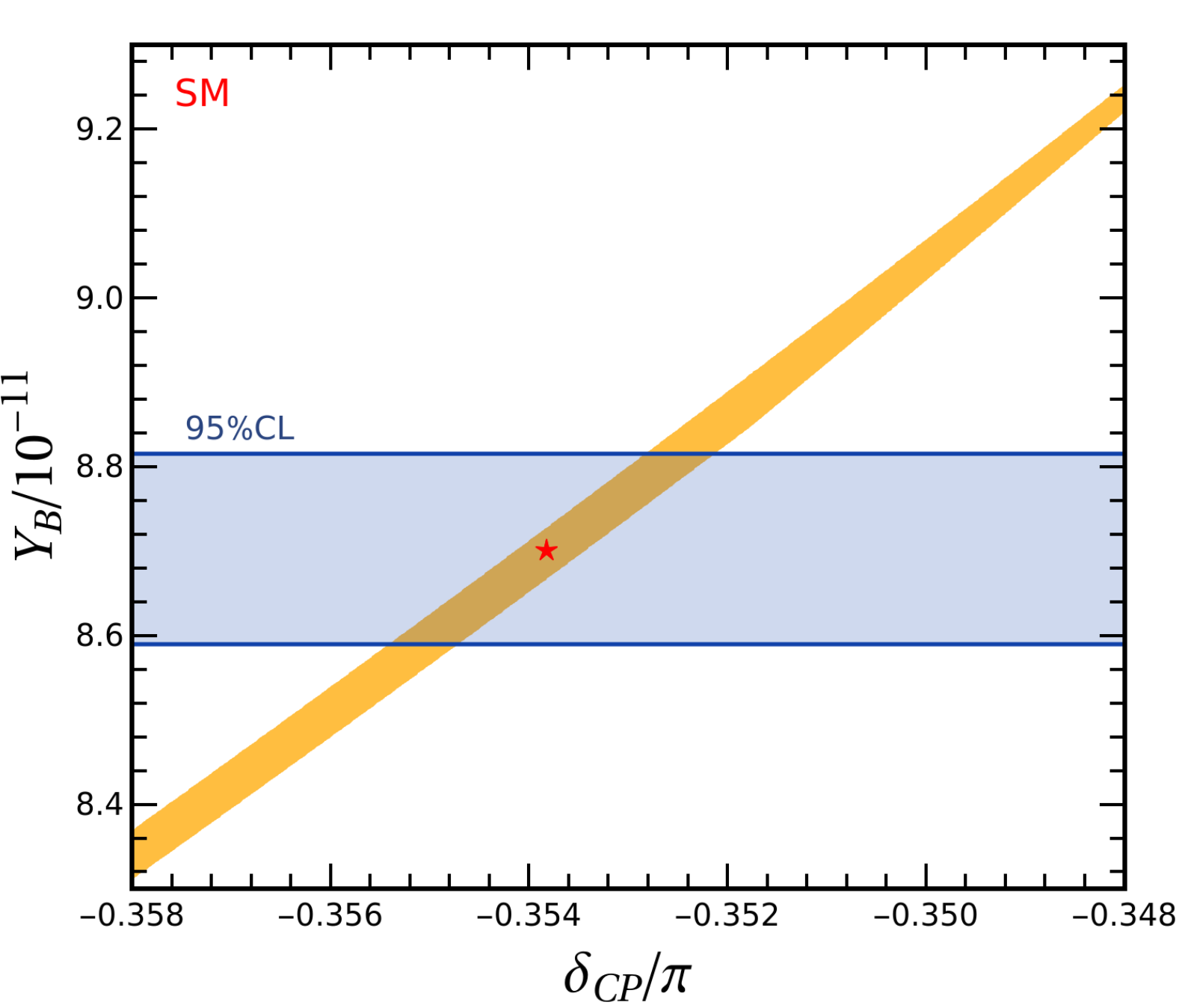}
\end{tabular}
\caption{\label{fig:SM_deltaCP_YB} The correlation between $Y_B$ and $\delta_{CP}$
for the new Littlest Seesaw model
in SM where $M_{1}=1.176\times 10^{11}$GeV. The Planck result for the baryon asymmetry $Y_B$ at $95\%$ CL is represented by the horizontal band~\cite{Aghanim:2018eyx}. The red star denotes the best fitting point at which the $\chi^2$ function reaches a global minimum.}
\end{figure}

In the present work, we shall discuss the flavoured thermal leptogenesis scenario in 2RHN model with hierarchical Majorana masses ($M_{1}\ll M_{2}$), where the two right-handed neutrino masses $M_{1}=x_{\text{atm}}\langle\xi_{\text{atm}}\rangle$ and $M_{2}=x_{\text{sol}}\langle\xi_{\text{sol}}\rangle$ are the masses of the right-handed neutrinos $N_{\text{atm}}$ and $N_{\text{sol}}$, respectively. The flavoured thermal leptogenesis has been studied in detail~\cite{Abada:2006fw,Abada:2006ea,Nardi:2006fx}. It was shown that the Boltzmann equations describing the asymmetries in flavour space are given by~\cite{Antusch:2006cw}
\begin{align}
\label{eq:Bol_Eq1}
& \frac{\text{d} Y_{N_{\text{atm}}}}{\text{d} z}&=& K\left(Y_{N_{\text{atm}}}^{\text{eq}}-Y_{N_{\text{atm}}}\right)\frac{z  f_{1}(z)K_{1}(z)}{K_{2}(z)}\,,\\
\label{eq:Bol_Eq2}
&\frac{\text{d} Y_{\Delta_{\alpha}}}{\text{d} z}&=&\varepsilon_{1 \alpha}^{\text{SM}}  K\left(Y_{N_{\text{atm}}}^{\text{eq}}-Y_{N_{\text{atm}}}\right)\frac{z  f_{1}(z)K_{1}(z)}{K_{2}(z)}+K_{\alpha}  Y_{N_{\text{atm}}}^{\text{eq}} \frac{z  f_{2}(z)K_{1}(z)}{K_{2}(z)}\frac{\sum_{\gamma} A_{\alpha \gamma}^{\text{SM}} Y_{\Delta_{\gamma}}}{Y_{\ell}^{\text{eq}}}\,.
\end{align}
There is no sum over $\alpha$ in the last term of Eq.~\eqref{eq:Bol_Eq2}, $z= M_1/T$ with $T$ being the temperature, $K_{1}(z)$ and $K_{2}(z)$ are the modified Bessel functions of the second kind, and $Y_{N_{\text{atm}}}$ denotes the density of the lightest right-handed neutrino $N_{\text{atm}}$\footnote{We find that the observed excess of matter over antimatter can not be generated in the new Littlest Seesaw model if $N_{\text{sol}}$ is the lightest right-handed neutrino.}. $Y_{N_{\text{atm}}}^{\text{eq}}$ and $Y_{\ell}^{\text{eq}}$ stand for the corresponding equilibrium number densities and they take the following form
\begin{equation}\label{eq:Yleq_Ynieq}
Y_{\ell}^{\text{eq}}\simeq\frac{45}{\pi^4g_{*}^{\text{SM}}},\qquad Y_{N_{\text{atm}}}^{\text{eq}}\simeq\frac{45z^{2}K_{2}(z)}{2\pi^4g_{*}^{\text{SM}}}\,,
\end{equation}
with $g_{*}^{\text{SM}}=106.75$. In order to obtain phenomenologically viable baryon asymmetry,  the lighter right-handed neutrino mass  $M_{1}$ is assumed in the interval of $10^9$ GeV $\leq M_{1}\leq10^{12}$ GeV. In this scenario, the $\tau$ Yukawa interaction is in equilibrium, the $e$ and $\mu$ flavours are indistinguishable, and the lepton number densities and $Y_{\Delta_\alpha}$ in the $e$ and $\mu$ flavour can be combined to $Y_{2}\equiv Y_{e+\mu}$ and $Y_{\Delta_2}\equiv Y_{\Delta_e+\Delta_\mu}$~\cite{Abada:2006ea,Abada:2006fw,Nardi:2006fx}. In this temperature range, the matrix $A^{\text{SM}}$ in the Boltzmann equation Eq.~\eqref{eq:Bol_Eq2} is given by~\cite{Abada:2006ea}
\begin{equation}\label{eq:Asm}
  A^{\text{SM}}=\frac{1}{589}\left( \begin{array}{cc}-417   &~ 120  \\ 30 &~ -390 \end{array}\right)\,,
\end{equation}
which arises from the washout term. The functions $f_1(z)$ and $f_2(z)$ in Eqs.~\eqref{eq:Bol_Eq1} and \eqref{eq:Bol_Eq2} account for the presence of $\Delta L=1$ scatterings and scatterings in the washout term of the asymmetry respectively~\cite{Giudice:2003jh,Buchmuller:2004nz}.
In the strong washout regime, $f_1(z)$ and $f_2(z)$ can be approximated as~\cite{Giudice:2003jh,Buchmuller:2004nz}
\begin{equation}
\label{eq:f1f2}
f_{1}(z)=2f_{2}(z)=\left[\frac{K_{s}}{zK }+\frac{z}{t} \ln \left(1+\frac{t}{z}\right)\right] \frac{K_{2}(z)}{K_{1}(z)}\,,
\end{equation}
with
\begin{equation}\label{eq:a-Ks-over-K}
t=\frac{K}{K_{s}\ln(M_{1}/M_{h})}, \qquad \quad
\frac{K_{s}}{K}=\frac{9}{8\pi^2}\,.
\end{equation}
where $M_{h}=125$ GeV is the mass of the Higgs boson. The flavoured CP asymmetries in the decays of the lightest RHN $N_{\text{atm}}$ into Higgs and leptons of different flavours are of the form~\cite{Covi:1996wh,Buchmuller:2004nz,Buchmuller:2005eh,Davidson:2008bu}
\begin{equation}\label{eq:e1a}
  \varepsilon_{1 \alpha}^{\text{SM}}
  =\frac{1}{8 \pi{(\lambda \lambda^\dagger)}_{11}}
  \left\{\Im\left[(\lambda \lambda^{\dagger})_{1 2}\lambda_{1 \alpha}\lambda^{*}_{2 \alpha}\right]g^{\text{SM}}(y)
  +\frac{1}{y-1} \Im\left[(\lambda\lambda^{\dagger})_{2 1}\lambda_{1 \alpha}\lambda^{*}_{2 \alpha}\right]\right\}\,,
\end{equation}
where $y=M^2_2/M^2_1$, $\lambda$ is the neutrino Yukawa coupling matrix and it is a $2\times3$ matrix with the following form
\begin{equation}
\lambda=\begin{pmatrix}\label{eq:nu_Yukawa}
0 &~    -a &~ a  \\
be^{-\frac{i\pi}{4}}   &~  \frac{5}{2}  b e^{-\frac{i\pi}{4}} &~  -\frac{1}{2}be^{-\frac{i\pi}{4}}\\
\end{pmatrix}\,,
\end{equation}
where $a=|y_{\text{atm}}v_{\text{atm}}|/v$, $b=|y_{\text{sol}}v_{\text{sol}}|/v$ and $v=246/\sqrt{2}$ GeV is the VEV of the Higgs field. The loop function $g^{\text{SM}}(y)$ in Eq.~\eqref{eq:e1a} can be written as
\begin{equation}
g^{\text{SM}}(y)=\sqrt{y}\left[\frac{1}{1-y}+1-(1+y) \ln \left(\frac{1+y}{y}\right)\right] \stackrel{y \gg 1}{\longrightarrow}-\frac{3}{2 \sqrt{y}}\,.
\end{equation}
Since hierarchical RHN masses $M_{1} \ll M_{2}$ ($y\gg1$) are assumed, we can get the following approximation formula for the decay asymmetry
\begin{equation}
\label{eq:e1a2}
\varepsilon_{1 \alpha}^{\text{SM}}=-\frac{3}{16 \pi} \frac{\Im\left[(\lambda \lambda^{\dagger})_{1 2}\lambda_{1 \alpha}\lambda^{*}_{2 \alpha}\right]}{{\left(\lambda \lambda^{\dagger}\right)}_{11}} \frac{M_{1}}{M_{2}} \,.
\end{equation}
For the breaking pattern discussed in section~\ref{sec:TDCP}, the flavour dependent decay asymmetries are:
\begin{equation}
\varepsilon_{1 e}^{\text{SM}}=0, \qquad \varepsilon_{1 \mu}^{\text{SM}}=\frac{3}{16 \pi}\frac{M_{1}}{M_{2}}(x-1)(x-2)b^{2}\sin\eta, \qquad
\varepsilon_{1\tau}^{\text{SM}}=\frac{3}{16 \pi}\frac{M_{1}}{M_{2}}x(x-1)b^{2}\sin\eta \,.
\end{equation}
In the new Littlest Seesaw model with $x=-1/2$, $\eta=-\pi/2$, $\varepsilon_{1 \alpha}^{\text{SM}}$ ($\alpha=e,\mu,\tau$) read as
\begin{equation}\label{eq:emt_e1e}
\varepsilon_{1 e}^{\text{SM}}=0, \qquad \varepsilon_{1 \mu}^{\text{SM}}=-\frac{45}{64 \pi}\frac{M_{1}}{M_{2}}b^{2}, \qquad
\varepsilon_{1\tau}^{\text{SM}}=-\frac{9}{64 \pi}\frac{M_{1}}{M_{2}}b^{2} \,.
\end{equation}
Note that $b^{2}/M_{2}\propto m_{s}$ which is defined in Eq.~\eqref{eq:mnu}, once the value of $m_{s}$ is fixed through the masses squared differences $\Delta m^2_{21}$ and $\Delta m^2_{31}$, $\varepsilon_{1\mu}^{\text{SM}}$ and $\varepsilon_{1\tau}^{\text{SM}}$ only depend on the lightest right-handed neutrino mass $M_{1}$. In addition to the decay asymmetry, the washout parameter $K_{\alpha}$, which appears in the washout term of the Boltzmann equation, is given by
\begin{equation}
  \label{eq:Ka}
  K_{\alpha}=\frac{\widetilde{m}_{1\alpha}}{m^{*}_{\text{SM}}},\qquad\,K=\sum_{\alpha}K_{\alpha}\,,
\end{equation}
where $m^{*}_{\text{SM}}\simeq1.08\times10^{-3}$ eV and the washout mass $\widetilde{m}_{1\alpha}$ parameterizes the decay rate of $N_\text{atm}$ into the leptons of flavour $\alpha$ with
\begin{equation}\label{eq:tildem1a}
\widetilde{m}_{1\alpha}\equiv\frac{|\lambda_{1\alpha}|^2v^2}{M_{1}}\,.
\end{equation}
From the Yukawa coupling matrix $\lambda$ given in Eq.~\eqref{eq:nu_Yukawa}, we find $K_{\alpha}$ is given by
\begin{equation}\label{eq:Kemt}
K_{e}=0, \qquad  K_{\mu}=K_{\tau}=\frac{m_{a}}{m^{*}_{\text{SM}}}\,.
\end{equation}
where $m_{a}=a^2v^2/M_{1}$ is defined in Eq.~\eqref{eq:mnu}. For the new Littlest Seesaw model, the best fitting value of $m_a$ is $m_a=22.366\text{meV}$\footnote{The parameter $m_a$ should be in the range $21.707\text{meV}\leq m_a\leq23.019\text{meV}$ in order to be compatible with present neutrino oscillation data.} as shown in table~\ref{tab:bf}. Then we can obtain the washout parameters $K_{\alpha}$ as follows
\begin{equation}
K_{e}+K_{\mu}=20.709\gg 1, \qquad  K_{\tau}=20.709\gg 1\,.
\end{equation}
Hence all flavours are in the strong washout region. Numerically solving the Boltzmann equations in Eqs.~(\ref{eq:Bol_Eq1}, \ref{eq:Bol_Eq2}), we find that the observed baryon asymmetry  $Y_{B}=8.7\times10^{-11}$ fix the lightest right-handed neutrino mass in the new Littlest Seesaw model:
\begin{equation}
M_{1}=1.176\times10^{11} \text{GeV}\,.
\end{equation}
We plot the baryon asymmetry $Y_{B}$ with respect to the Dirac CP phase $\delta_{CP}$ in figure~\ref{fig:SM_deltaCP_YB}. The width of the line comes from varying $m_a$ and $r$ over their allowed ranges, where all three mixing angles and two neutrino mass squared differences are required to lie in the experimentally preferred $3\sigma$ ranges~\cite{Esteban:2018azc}.

\subsection{\label{sec:lep_MSSM} Leptogenesis for the new Littlest Seesaw model in the MSSM}

\begin{figure}[t!]
\centering
\begin{tabular}{c}
\includegraphics[width=0.60\linewidth]{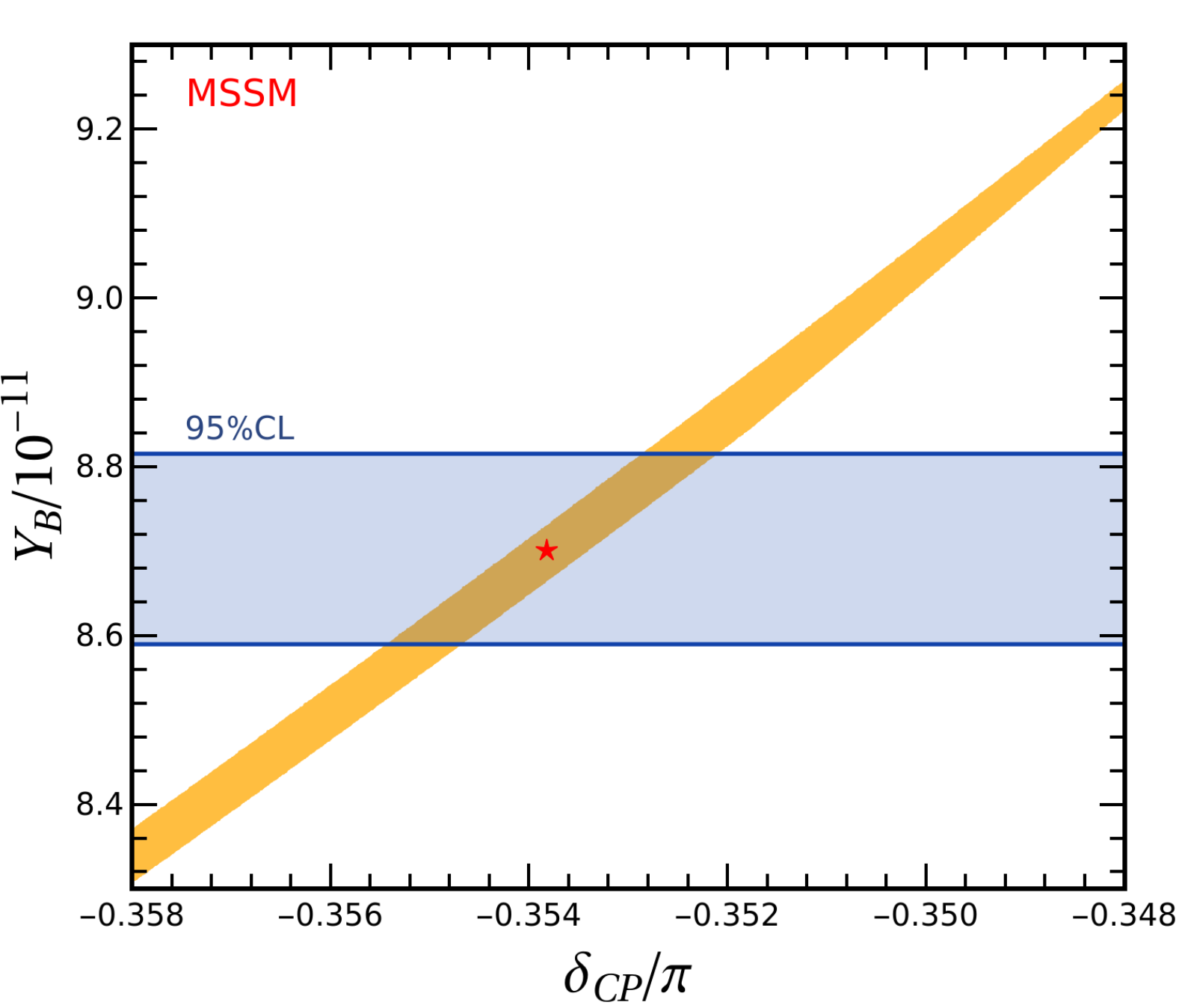}
\end{tabular}
\caption{\label{fig:MSSM_deltaCP_YB}
The correlation between $Y_B$ and $\delta_{CP}$ for the new Littlest Seesaw model in the MSSM where we take $M_{1}=3.992\times 10^{10}$GeV. The Planck result for the baryon asymmetry $Y_B$ at $95\%$ CL is represented by the horizontal band~\cite{Aghanim:2018eyx}. The red star denotes the best fitting point at which the $\chi^2$ function reaches a global minimum. }
\end{figure}

In the MSSM, the final baryon asymmetry can be computed from the following formula~\cite{Harvey:1990qw}
\begin{equation}\label{eq:YB MSSM}
Y_{B}=\frac{10}{31}\sum_{\alpha} \hat{Y}_{\Delta_\alpha}\,.
\end{equation}
In the MSSM, the contributions of $\widetilde {N}_1$ and $\widetilde {L}_\alpha$ should be considered, which are the supersymmetric partners of the lightest right-handed neutrino $N_1$ and the lepton doublet $L_\alpha$ respectively. In other words, the densities $Y_{\widetilde N_1}$ and $Y_{\widetilde \alpha}$ should be included in the Boltzmann equations. Then the Boltzmann equations in MSSM are given by~\cite{Antusch:2006cw}
\begin{eqnarray}
\nonumber \frac{\text{d} Y_{N_\text{atm}}}{\text{d} z} &=&2 K(Y^\text{eq}_{N_\text{atm}}-Y_{N_\text{atm}} )\frac{z  f_1 (z)K_1 (z)}{K_2 (z)} \, ,\\
\nonumber\frac{\text{d} Y_{\widetilde N_\text{atm}}}{\text{d} z} &=& 2 K(Y^\text{eq}_{\widetilde N_\text{atm}}-Y_{\widetilde N_\text{atm}} )\frac{z f_1 (z)K_1 (z)}{K_2 (z)}\, ,\\
\nonumber   \frac{\text{d} \hat Y_{\Delta_\alpha}}{\text{d} z} &=&K(\varepsilon_{1\alpha}^{\text{MSSM}}+\varepsilon_{1\widetilde \alpha}^{\text{MSSM}}) (Y^\text{eq}_{N_\text{atm}}-Y_{N_\text{atm}} )\frac{z f_1 (z)K_1 (z)}{K_2 (z)} \\
\nonumber &&+K(\varepsilon_{\widetilde 1\alpha}^{\text{MSSM}}+\varepsilon_{\widetilde 1\widetilde \alpha}^{\text{MSSM}}) (Y^\text{eq}_{\widetilde N_\text{atm}}-Y_{\widetilde N_\text{atm}} )\frac{z f_1 (z)K_1 (z)}{K_2 (z)}
\nonumber \\
\label{eq:MSSM_be}&&+
K_{\alpha}(Y^\text{eq}_{N_\text{atm}}+Y^\text{eq}_{\widetilde N_\text{atm}})    \frac{zf_2 (z)K_1 (z)}{K_2 (z)}
 \frac{\sum_\gamma  A_{\alpha\gamma}^{\text{MSSM}} \hat Y_{\Delta_\gamma}}{\hat Y^\text{eq}_{\ell}} \,,
\end{eqnarray}
where the total (particle and sparticle) $B/3 - L_\alpha$ asymmetries denoted as $\hat Y_{\Delta_\alpha}$ and
\begin{equation}
\hat{Y}_{\ell}^{\text{eq}}=Y^{\text{eq}}_{\tilde{\ell}}+Y^{\text{eq}}_{\ell},\qquad
Y^{\text{eq}}_{\tilde{\ell}}\simeq Y^{\text{eq}}_{\ell}\simeq\frac{45}{\pi^4g_{*}^{\text{MSSM}}} ,\qquad Y_{N_\text{atm}}^{\text{eq}}=Y_{\widetilde N_\text{atm}}^{\text{eq}}\simeq\frac{45z^{2}K_{2}(z)}{2\pi^4g_{*}^{\text{MSSM}}}\,,
\end{equation}
with $g_{*}^{\text{MSSM}}=228.75$. The matrix $A^{\text{MSSM}}$ in Eq.~\eqref{eq:MSSM_be} depends on which MSSM interactions are in thermal equilibrium at the temperatures where leptogenesis takes place. Here we shall consider the case that the lightest right-handed neutrino mass $M_{1}$ is between $(1+\tan^2\beta)\times10^{9}$ GeV and $(1+\tan^2\beta)\times10^{12}$ GeV, where only the $\tau$ Yukawa couplings are in thermal equilibrium. Then the relevant flavour-dependent asymmetries are $\hat Y_{\Delta_2} \equiv \hat Y_{\Delta_e+\Delta_\mu}$ and $\hat Y_{\Delta_\tau}$, and $A^{\text{MSSM}}$ is given by
\begin{equation}\label{eq:Amssm}
A^{\text{MSSM}}=\frac{1}{761}\left( \begin{array}{rr}-541   &~ 152  \\ 46  &~ -494 \end{array}\right)\,.
\end{equation}
In the MSSM, the decay asymmetries are all equal ($\varepsilon^\text{MSSM}_{1\alpha} =
\varepsilon^\text{MSSM}_{1\widetilde \alpha} =
\varepsilon^\text{MSSM}_{\widetilde 1\alpha} =
\varepsilon^\text{MSSM}_{\widetilde 1\widetilde \alpha}$)~\cite{Covi:1996wh}. As a consequence, we will only show the results of $\varepsilon^\text{MSSM}_{1\alpha}$ in the following. Under the assumption of $M_{1} \ll M_{2}$, the CP asymmetry $\varepsilon^\text{MSSM}_{1\alpha}$ ($\alpha=e,\mu,\tau$) in the MSSM is given by
\begin{equation}
\label{eq:MSSM_e1a}
\varepsilon_{1 \alpha}^{\text{MSSM}}=\frac{1}{8 \pi{\left(\lambda \lambda^{\dagger}\right)}_{11}}\Im\left[(\lambda \lambda^{\dagger})_{1 2}\lambda_{1 \alpha}\lambda^{*}_{2 \alpha}\right]g^{\text{MSSM}}\left(\frac{M_{2}^{2}}{M_{1}^{2}}\right)\,,
\end{equation}
where the function $g^{\text{MSSM}}(y)$ is of the following form
\begin{equation}\label{eq:gmssm}
g^{\text{MSSM}}(y)=\sqrt{y}\left[\frac{2}{1-y}-\ln \left(\frac{1+y}{y}\right)\right] \stackrel{y \gg 1}{\longrightarrow}-\frac{3}{\sqrt{y}}\,.
\end{equation}
Inserting the expression of function $ g^{\text{MSSM}}(M^2_2/M^2_1)$ into $\varepsilon_{1 \alpha}^{\text{MSSM}}$ in Eq.~\eqref{eq:MSSM_e1a} we find
\begin{equation}
\varepsilon_{1 e}^{\text{MSSM}}=0, \quad \varepsilon_{1 \mu}^{\text{MSSM}}=\frac{3}{8 \pi}\frac{M_{1}}{M_{2}}(x-1)(x-2)b^{2}\sin\eta, \quad
\varepsilon_{1\tau}^{\text{MSSM}}=\frac{3}{8 \pi}\frac{M_{1}}{M_{2}}x(x-1)b^{2}\sin\eta\,,
\end{equation}
for the most general case.  In the new Littlest Seesaw model with $x=-1/2$, $\eta=-\pi/2$, the flavour dependent decay asymmetries are determined to be,
\begin{equation}\label{eq:MSSMe1e}
\varepsilon_{1 e}^{\text{MSSM}}=0, \qquad \varepsilon_{1 \mu}^{\text{MSSM}}=-\frac{45}{32 \pi}\frac{M_{1}}{M_{2}}b^{2}, \qquad  \varepsilon_{1 \tau}^{\text{MSSM}}=-\frac{9}{32 \pi}\frac{M_{1}}{M_{2}}b^{2}\,.
\end{equation}
The washout parameters $K_{\alpha}$ and $K$ in Eq.~\eqref{eq:MSSM_be} are defined as
\begin{equation}\label{eq:MSSM_washout}
K_{\alpha}=\frac{\tilde{m}_{1\alpha}}{m^{*}_{\text{MSSM}}},\qquad \tilde{m}_{1\alpha}\equiv\frac{|\lambda_{1\alpha}|^2v_{u}^2}{M_{1}}, \qquad K=\sum_{\alpha}K_{\alpha}  \,,
\end{equation}
with
\begin{equation}
v_{u}=v\sin\beta, \qquad  m^{*}_{\text{MSSM}}\simeq\sin^2\beta\times1.58\times10^{-3}\,\text{eV}\,.
\end{equation}
The expressions of the washout parameters for the new Littlest Seesaw model are
\begin{equation}\label{eq:Kemt_mssm}
K_{e}=0, \qquad  K_{\mu}=K_{\tau}=\frac{m_{a}}{m^{*}_{\text{MSSM}}}\,,
\end{equation}
with $m_a=a^2v^2_u/M_{1}$. At the best fitting of our model, the values of the washout parameters are
\begin{equation}
K_{e}+K_{\mu}=14.722\gg 1,\qquad K_{\tau}=14.722\gg 1\,,
\end{equation}
which implies all flavours are in the strong washout region. For illustration, we take $\tan{\beta}=5$ and we find the experimentally observed values of the baryon asymmetry can be obtained if the lightest right-handed neutrino mass
in the new Littlest Seesaw model is
\begin{equation}\label{eq:MSSM_M1}
M_{1}=3.992\times10^{10} \text{GeV}\,.
\end{equation}
The correlation between $Y_B$ and $\delta_{CP}$
in the new Littlest Seesaw model is displayed in figure~\ref{fig:MSSM_deltaCP_YB}.

\section{\label{sec:model}  Explicit model for the new Littlest Seesaw}

As we have shown in previous sections, the new Littlest Seesaw model can describe the experiment data of lepton mixing angles, neutrino masses and matter-antimatter asymmetry of the Universe very well. In this section, we shall construct an explicit model based on the model independent analysis of section~\ref{sec:TDCP}. The vacuum alignments $\langle\phi_{\text{atm}}\rangle\propto \left(0,1,-1\right)$, $\langle\phi_{\text{sol}}\rangle\propto \left(1,-1/2,5/2\right)$ and the phase parameter $\eta=-\pi/2$ will be naturally realized in our model. We impose the $S_4$ flavour symmetry as well as CP symmetry. The standard supersymmetric driving field mechanism~\cite{Altarelli:2005yx} which we adopt in our model requires a $U(1)_R$ symmetry related to the usual $R$-parity. Furthermore, we also introduce the shaping symmetry $Z_{5}\times Z_8$ which allows us to forbid unwanted terms and achieve the desired vacuum alignment. The auxiliary symmetry $Z_{8}$ is helpful to generate the phase $\eta=-\pi/2$. The shaping symmetry $Z_{5}$ requires the electron, muon and tauon mass terms couple with different powers of flavon fields. Hence $Z_{5}$ helps to reproduce the observed charged lepton mass hierarchies. Here we choose the right-handed charged leptons as $S_4$ singlets, where $e^{c}$ and $\tau^{c}$ transform as $\mathbf{1}$ while $\mu^{c}$  transforms as $\mathbf{1^\prime}$. The three generations of left-handed lepton doublets $L$ are unified to an $S_4$ triplet $\mathbf{3}$. The two right-handed neutrinos $\nu^{c}_{\text{atm}}$ and $\nu^{c}_{\text{sol}}$ are assigned to $\mathbf{1}$ and $\mathbf{1}'$ of $S_4$, respectively. The field content and their classification under the flavour symmetry $S_4\times Z_5\times Z_8$ are listed in table~\ref{tab:tran_pro}. The driving fields are indicated with the superscript ``0'' and they carry two units of $R$ charge, both flavon fields and Higgs are uncharged under $U(1)_R$, and the $R$-charge of the matter fields is equal to one. Since both flavon fields and driving fields are Standard Model singlets, our model is anomaly free under the Standard Model gauge transformation. As regards possible discrete anomalies, in principle they may be cancelled by adding extra states under the discrete group, but this is beyond the scope of this paper. In the following, we first discuss the vacuum alignment of the model, then specify the structure of the model.

\begin{table}[t!]
\renewcommand{\tabcolsep}{0.5mm}
\begin{center}
\begin{tabular}{|c|c|c|c|c|c|c|c||c|c||c|c|c|c||c|c|c||c|c|c||c|c|c||c|c|c|c||c|c||c|c|c|c|c|c|c|c|c|c|c|c|}\hline\hline
& $L$ & $e^c$  &  $\mu^c$ &  $\tau^c$  & $\nu^{c}_{\text{atm}}$ & $\nu^{c}_{\text{sol}}$  &$H_{u,d}$ &  $\chi_{l}$ &  $\phi_{l}$   & $\xi_{a}$ & $\zeta_{a}$ &    $\eta_{a}$ & $\phi_{a}$  & $\xi_{s}$ &   $\varphi_{s}$ &  $\phi_{s}$ &  $\eta^0_{l}$ &    $\xi^0_{l}$  & $\zeta^0_{l}$  & $\xi^0_{a}$ &  $\eta^0_{a}$ & $\phi^0_{a}$  & $\xi^0_{s}$ &  $\varphi^0_s$ &  $\zeta^0_s$ & $\phi^0_s$ & $\sigma^0_{1,2}$  \\ \hline

$S_4$  & $\mathbf{3}$ &  $\mathbf{1}$  &  $\mathbf{1^\prime}$ & $\mathbf{1}$  & $\mathbf{1}$ & $\mathbf{1^\prime}$ & $\mathbf{1}$  &  $\mathbf{3}$ &  $\mathbf{3}$  &  $\mathbf{1}$ & $\mathbf{1^\prime}$ &   $\mathbf{2}$ & $\mathbf{3}$  & $\mathbf{1}$  & $\mathbf{3}$  &  $\mathbf{3}^{\prime}$ &  $\mathbf{2}$  & $\mathbf{1}$  &  $\mathbf{1}$ &  $\mathbf{1}$  & $\mathbf{2}$  &  $\mathbf{3^\prime}$ &  $\mathbf{1}$  &    $\mathbf{3^\prime}$ &  $\mathbf{1^\prime}$ &  $\mathbf{3^\prime}$  &  $\mathbf{1}$ \\

$Z_5$  & $1$ &  $\omega^2_5$  &  $\omega^3_5$  & $\omega^4_5$  & $1$ & $1$ & $1$ &
 $\omega^4_5$  & $\omega_5$  &  $1$  & $\omega^4_5$  & $1$  & $1$   & $1$  &   $\omega_5$   & $1$ &
 $\omega^2_5$ & $\omega^3_5$ &  $1$  & $\omega_5$   & $1$  & $\omega_5$ & $1$ &   $\omega^4_5$   & $\omega^4_5$ & $\omega^4_5$ & $1$    \\

$Z_8$ & $1$ & $\omega^3_8$  & $i$ & $\omega_8$ & $-i$  &  $\omega^5_8$   & $1$ &
$i$  & $\omega^7_8$   &  $-1$  & $-1$  & $-1$  & $i$   & $-i$  &   $-i$   & $\omega^3_8$ &
 $-1$ & $i$ &  $\omega^7_8$  & $-1$   & $-1$  & $i$ & $i$ &   $1$   & $\omega^7_8$ & $\omega^7_8$   & $1$  \\  \hline\hline


\end{tabular}
\caption{\label{tab:tran_pro} The matter field, flavon fields, driving fields and their transformation properties under the flavour symmetry $S_4\times  Z_5\times Z_8$ in model, where $\omega_5=e^{2\pi i/5}$ and $\omega_8=e^{\pi i/4}$. }
\end{center}
\end{table}


\subsection{\label{subsec:alignment}Vacuum alignment}

We employ the now-standard $F$-term alignment mechanism to arrange the vacuum~\cite{Altarelli:2005yx} in our model. It requires that all terms in the superpotential must carry two units of $R$ charge. Therefore each term in the superpotential contains either two matter superfields or only one driving field. The minimum of the scalar potential is determined by vanishing $F$-terms of the driving fields. The leading order (LO) driving superpotential $w_{d}$ in which each term contains one driving field invariant under $S_4\times Z_{5}\times Z_8$ can be written as
\begin{eqnarray}
\nonumber w_{d}&=&f_1\left(\eta^0_l\left(\chi_l\chi_l\right)_{\mathbf{2}}\right)_{\mathbf{1}}+f_2\xi^0_{l}\left(\phi_l\phi_l\right)_{\mathbf{1}}
+f_3\zeta^0_{l}\left(\phi_l\chi_l\right)_{\mathbf{1}}+f_4\xi^0_{a}\left(\chi_l\phi_a\right)_{\mathbf{1}}+M_{\eta}\left(\eta^0_{a}\eta_a\right)_{\mathbf{1}} \\
\nonumber &&+f_5\left(\eta^0_{a}\left(\phi_a\phi_a\right)_{\mathbf{2}}\right)_{\mathbf{1}}+f_6\zeta_a\left(\phi^0_{a}\phi_{a}\right)_{\mathbf{1^\prime}}
+f_7\left(\phi^0_{a}\left(\eta_a\chi_l\right)_{\mathbf{3^\prime}}\right)_{\mathbf{1}}
+M^2_{\sigma_1}\sigma^0_1+f_{8}\sigma^0_1\xi^2_a \\
\nonumber &&+M^2_{\sigma_2}\sigma^0_2+f_{9}\sigma^0_2\left(\eta_a\eta_a\right)_{\mathbf{1}}
+f_{10}\left(\varphi^0_{s}\left(\phi_a\varphi_s\right)_{\mathbf{3^\prime}}\right)_{\mathbf{1}}
+f_{11}\zeta^0_s\left(\varphi_s\phi_s\right)_{\mathbf{1^\prime}}+f_{12}\left(\phi^0_{s}\left(\phi_l\phi_a\right)_{\mathbf{3^\prime}}\right)_{\mathbf{1}}\\
&&+f_{13}\left(\phi^0_{s}\left(\varphi_s\phi_s\right)_{\mathbf{3^\prime}}\right)_{\mathbf{1}}+M_{\xi}\xi^0_s\xi_s +f_{14}\xi^0_s\left(\phi_s\phi_s\right)_{\mathbf{1}}\,,
\end{eqnarray}
where $(\ldots)_{\mathbf{r}}$ stands for a contraction into the $S_4$ irreducible representation $\mathbf{r}$. Because we impose CP as symmetry on the model, all the couplings $f_i$ ($i=1,\cdots,14$) and mass parameters $M_{\eta}$, $M_{\xi}$, $M_{\sigma_1}$, $M_{\sigma_2}$ are constrained to be real. The VEVs of the flavon $\chi_l$ can be obtained from the vanishing of the derivatives of $w_d$ with respect to each component of the driving fields $\eta^0_{l}$, i.e.
\begin{eqnarray}
\label{eq:align_ch}
\nonumber&&\frac{\partial w_{d}}{\partial\eta^{0}_{l_{1}}}=f_{1} \left(2 \chi_{l_1} \chi_{l_2}+\chi_{l_3}^2\right)=0\,,\\
&&\frac{\partial w_{d}}{\partial\eta^{0}_{l_{2}}}=f_{1} \left(2 \chi_{l_1} \chi_{l_3}+\chi_{l_2}^2\right)=0\,.
\end{eqnarray}
One solution to these equations is
\begin{equation}\label{eq:vev_chil}
\langle\chi_{l}\rangle=v_{\chi_{l}}(1,0,0)^T\,,
\end{equation}
where $v_{\chi_{l}}$ is undetermined. In the charged lepton sector, the $F$-term conditions of the driving fields $\xi^0_{l}$ and $\zeta^0_{l}$ give the vacuum alignment of $\phi_l$,
\begin{eqnarray}
\nonumber&&\frac{\partial w_{d}}{\partial\xi^{0}_l}=f_{2} \left(\phi_{l_1}^2+2 \phi_{l_2} \phi_{l_3}\right)=0\,,\\
&&\frac{\partial w_{d}}{\partial\zeta^{0}_l}=f_{3} (\chi_{l_1} \phi_{l_1}+\chi_{l_2} \phi_{l_3}+\chi_{l_3} \phi_{l_2})=0\,.
\end{eqnarray}
Given the vacuum of $\chi_l$ in Eq.~\eqref{eq:vev_chil}, we find the alignment of $\phi_{l}$ is
\begin{equation}\label{eq:vev_phil}
\langle\phi_{l}\rangle=v_{\phi_{l}}\left(0, 1,0\right)^T\,,
\end{equation}
with $v_{\phi_{l}}$ undetermined. In the atmospheric neutrino sector, the $F$-term conditions associated with the driving fields $\xi^0_{a}$, $\eta^0_{a}$ and $\phi^{0}_{a}$ read
\begin{eqnarray}
\nonumber&&\frac{\partial w_{d}}{\partial\xi^0_{a}}=f_{4} (\chi_{l_1} \phi_{a_1}+\chi_{l_2} \phi_{a_3}+\chi_{l_3} \phi_{a_2})=0\,, \\
\nonumber&&\frac{\partial w_{d}}{\partial\eta^{0}_{a_{1}}}=M_{\eta} \eta_{a_2}+f_{5} \left(2 \phi_{a_1} \phi_{a_2}+\phi_{a_3}^2\right)=0\,,\\
\nonumber&&\frac{\partial w_{d}}{\partial\eta^{0}_{a_{2}}}=M_{\eta} \eta_{a_1}+f_{5} \left(2 \phi_{a_1} \phi_{a_3}+\phi_{a_2}^2\right)=0\,,\\
\nonumber&&\frac{\partial w_{d}}{\partial\phi^{0}_{a_{1}}}=f_{6} \zeta_{a} \phi_{a_1}+f_{7} (\eta_{a_1} \chi_{l_2}-\eta_{a_2} \chi_{l_3})=0\,,\\
\nonumber&&\frac{\partial w_{d}}{\partial\phi^{0}_{a_{2}}}=f_{6} \zeta_{a} \phi_{a_3}+f_{7} (\eta_{a_1} \chi_{l_1}-\eta_{a_2} \chi_{l_2})=0\,,\\
&&\frac{\partial w_{d}}{\partial\phi^{0}_{a_{3}}}=f_{6} \zeta_{a} \phi_{a_2}+f_{7} (\eta_{a_1} \chi_{l_3}-\eta_{a_2} \chi_{l_1})=0\,.
\end{eqnarray}
A straightforward calculation shows that the vacuum expectation values of $\xi_{a}$, $\eta_{a}$ and $\phi_{a}$ are
\begin{equation}\label{eq:vev_atm}
 \langle\zeta_{a}\rangle=v_{\zeta_{a}}, \qquad \langle\eta_{a}\rangle=v_{\eta_{a}}\left(1, 1\right)^T,
\qquad \langle\phi_{a}\rangle=v_{\phi_{a}}\left(0, 1, -1\right)^T\,,
\end{equation}
with
\begin{equation}\label{eq:atm_cor}
v^2_{\phi_{a}} =-\frac{M_{\eta}}{f_5}v_{\eta_{a}}, \qquad
v_{\zeta_a}=-\frac{f_5f_7v_{\phi_a}v_{\chi_l}}{f_6M_{\eta}}\,.
\end{equation}
It is easy to check that the vacuum alignments of flavons $\eta_a$ and $\phi_{a}$ preserve the subgroup $Z^{U}_2$. Now we consider the phases of $v_{\phi_{a}}$ and $v_{\xi_a}$ which is the VEV of $\xi_a$. They are related by the $F$-flatness of $\sigma^{0}_{1,2}$:
\begin{eqnarray}\label{eq:xia_pha}
\nonumber &&\frac{\partial w_{d}}{\partial\sigma^{0}_{1}}=M_{\sigma_1}^2+f_{8} \xi_{a}^2=0\,, \\
&&\frac{\partial w_{d}}{\partial\sigma^{0}_{2}}=M_{\sigma_2}^2+2 f_{9} \eta_{a_1} \eta_{a_2}=0\,.
\end{eqnarray}
From Eqs.~\eqref{eq:atm_cor} and \eqref{eq:xia_pha}, we find
\begin{equation}
\frac{v_{\xi_a}}{v^2_{\phi_{a}}}=\frac{f_{5}M_{\sigma_1}}{M_{\sigma_2} M_{\eta}}\left(\frac{2f_{9}}{f_{8}}\right)^{1/2}\,.
\end{equation}
As all parameters in the right-hand side of above equation are real, consequently the phase of $\frac{v_{\xi_a}}{v^2_{\phi_{a}}}$ is $0$, $\pi$ or $\pm\pi/2$ for the product $f_{8}f_{9}>0$ or $f_{8}f_{9}<0$, respectively. The auxiliary symmetry $Z_8$ has played a critical role in generating the discrete possible values $0$, $\pi$, $\pm\pi/2$ for the phase of $v_{\xi_a}/v^2_{\phi_{a}}$. Subsequently we turn to discuss the vacuum alignment of the solar neutrino sector. The $F$-flatness condition of the driving field $\varphi^0_s$ gives
\begin{eqnarray}
\nonumber&&\frac{\partial w_{d}}{\partial\varphi^{0}_{s_1}}=f_{10} (2 \varphi_{s_1} \phi_{a_1}-\varphi_{s_2} \phi_{a_3}-\varphi_{s_3} \phi_{a_2})=0\,,\\
\nonumber&&\frac{\partial w_{d}}{\partial\varphi^{0}_{s_2}}=f_{10} (2 \varphi_{s_2} \phi_{a_2}-\varphi_{s_1} \phi_{a_3}-\varphi_{s_3} \phi_{a_1})=0\,,\\
&&\frac{\partial w_{d}}{\partial\varphi^{0}_{s_3}}=f_{10} (2 \varphi_{s_3} \phi_{a_3}-\varphi_{s_1} \phi_{a_2}-\varphi_{s_2} \phi_{a_1})=0\,,
\end{eqnarray}
which lead to the vacuum
\begin{equation}\label{eq:vev_varphis}
\langle\varphi_{s}\rangle=v_{\varphi_{s}}\left(2, -1,  -1\right)^T \,.
\end{equation}
The equations giving the vacuum structure for the flavon field $\phi_{s}$ are:
\begin{eqnarray}
\nonumber&&\frac{\partial w_{d}}{\partial\zeta^{0}_{s}}=f_{11} (\varphi_{s_1} \phi_{s_1}+\varphi_{s_2} \phi_{s_3}+\varphi_{s_3} \phi_{s_2})=0\,,\\
\nonumber&&\frac{\partial w_{d}}{\partial\phi^{0}_{s_{1}}}=f_{12} (2 \phi_{a_1} \phi_{l_1}-\phi_{a_2} \phi_{l_3}-\phi_{a_3} \phi_{l_2})+f_{13} (\varphi_{s_2} \phi_{s_3}-\varphi_{s_3} \phi_{s_2})=0\,,\\
\nonumber&&\frac{\partial w_{d}}{\partial\phi^{0}_{s_{2}}}=f_{12} (2 \phi_{a_2} \phi_{l_2}-\phi_{a_1} \phi_{l_3}-\phi_{a_3} \phi_{l_1})+f_{13} (\varphi_{s_3} \phi_{s_1}-\varphi_{s_1} \phi_{s_3})=0\,,\\
&&\frac{\partial w_{d}}{\partial\phi^{0}_{s_{3}}}=f_{12} (2 \phi_{a_3} \phi_{l_3}-\phi_{a_1} \phi_{l_2}-\phi_{a_2} \phi_{l_1})+f_{13} (\varphi_{s_1} \phi_{s_2}-\varphi_{s_2} \phi_{s_1})=0\,,
\end{eqnarray}
which uniquely determine the solar alignment,
\begin{equation}
\label{eq:vev_phis}
\langle\phi_{s}\rangle=v_{\phi_{s}}\left(1, -1/2,  5/2\right)^T \,, \quad \text{with} \quad
v_{\phi_{s}}=\frac{f_{12}v_{\phi_l}v_{\phi_a}}{3f_{13}v_{\varphi_{s}}}\,.
\end{equation}
We find the vacuum configurations of $\varphi_{s}$ and $\phi_{s}$ are invariant under the action of the subgroup $Z^{SU}_2$. Finally the $F-$term condition of $\xi^0_s$ is
\begin{equation}
\frac{\partial w_{d}}{\partial\xi^{0}_s}=M_{\xi} \xi_{s}+f_{14} \left(\phi_{s_1}^2+2 \phi_{s_2} \phi_{s_3}\right)=0\,,
\end{equation}
which  leads to the following relations
\begin{equation}\label{eq:pha_cor}
\frac{v^2_{\phi_s}}{v_{\xi_{s}}}=\frac{2M_{\xi}}{3f_{14}}\,.
\end{equation}
The phase parameter $\eta$ is exactly the phase of the ratio $\frac{v^2_{\phi_{s}}v_{\xi_{a}}}{v^2_{\phi_{a}}v_{\xi_{s}}}$ in our model. Form Eqs.~\eqref{eq:atm_cor} and \eqref{eq:pha_cor}, it is easy to obtain
\begin{equation}\label{eq:vev_ratio}
\frac{v^2_{\phi_{s}}v_{\xi_{a}}}{v^2_{\phi_{a}}v_{\xi_{s}}}=\frac{2f_{5}M_{\sigma_1}M_{\xi}}
{3f_{14}M_{\sigma_2}M_{\eta}}\left(\frac{2f_{9}}{f_{8}}\right)^{1/2}\,.
\end{equation}
All couplings and mass parameters in above equation are real due to CP symmetry, then we see the phase of the ratio $\frac{v^2_{\phi_{s}}v_{\xi_{a}}}{v^2_{\phi_{a}}v_{\xi_{s}}}$ is $e^{\frac{ik\pi}{2}}(i=0,1,...,3)$. In the present work we shall take the following solution
\begin{equation}\label{eq:vev_ratio_arg}
\text{arg}\left(\frac{v^2_{\phi_{s}}v_{\xi_{a}}}{v^2_{\phi_{a}}v_{\xi_{s}}}\right)=-\frac{\pi}{2}\,,
\end{equation}
which would happen for $f_{8}f_{9}<0$. Thus the desired vacuum alignment $\langle\phi_{a}\rangle\propto\left(0, 1, -1\right)^{T}$, $\langle\phi_{s}\rangle\propto\left(1,-1/2,5/2\right)^{T}$ and the phase $\eta=-\pi/2$ have been dynamically realized. In the following section we will find that the observed hierarchy among the charged lepton masses can be produced for
\begin{equation}
\frac{v_{\phi_l}}{\Lambda}\sim\lambda^2_C\,,
\end{equation}
where $\Lambda$  is the cut-off scale of the theory and $\lambda_C$ is  the Cabibbo angle with $\lambda_C\simeq0.23$. As usual, we expect that all the VEVs of flavons are of the same order of magnitude, i.e.
\begin{equation}
\frac{v_{\xi_a}}{\Lambda}\sim\frac{v_{\phi_a}}{\Lambda}\sim\frac{v_{\xi_s}}{\Lambda}\sim\frac{v_{\phi_s}}{\Lambda}\sim
\frac{v_{\eta_a}}{\Lambda}\sim\frac{v_{\zeta_a}}{\Lambda}\sim\frac{v_{\varphi_s}}{\Lambda}\sim\frac{v_{\chi_l}}{\Lambda}\sim\lambda^2_C\,.
\end{equation}
Successful leptogenesis fixes the atmospheric neutrino mass to be $3.992\times 10^{10}$ GeV (see Eq.~\eqref{eq:MSSM_M1}) which is of the same order as the flavon VEVs. Thus the cut-off scale $\Lambda$ is expected to be of order $10^{12}$ GeV. The next-to-leading-order (NLO) corrections to the flavon superpotential $w_d$ involve three flavon fields. When the NLO corrections are included, the original symmetry $S_4\rtimes H_{CP}$ is broken completely in the charged lepton, atmospheric neutrino and solar neutrino sectors. The NLO corrections to VEVs of all flavons are found to be suppressed by $\Phi/\Lambda\sim\lambda^2_C$ with respect to the LO contributions and therefore can be negligible, where $\Phi$ denotes any flavour fields.

\subsection{\label{subsec:model}The structure of the model}

The most relevant operators for charged lepton masses are given by
\begin{eqnarray}
\hskip-0.25in \nonumber  w_l&&=\frac{y_{\tau}}{\Lambda}\left(L\phi_{l}\right)_{\mathbf{1}}\tau^{c}H_d
+\frac{y_{\mu}}{\Lambda^2} \left(L\left(\phi_{l}\phi_{l}\right)_{\mathbf{3^\prime}}\right)_{\mathbf{1^\prime}}\mu^{c}H_d
 +\frac{y_{e_1}}{\Lambda^3}\left(L\phi_{l}\right)_{\mathbf{1}}\left(\phi_{l}\phi_{l}\right)_{\mathbf{1}}e^{c}H_d \\
\hskip-0.25in \label{eq:ch_Yukawa}&&+\frac{y_{e_2}}{\Lambda^3}\left(\left(L\phi_{l}\right)_{\mathbf{2}}\left(\phi_{l}\phi_{l}\right)_{\mathbf{2}}\right)_{\mathbf{1}}e^cH_d
+\frac{y_{e_3}}{\Lambda^3}\left(\left(L\phi_{l}\right)_{\mathbf{3}}\left(\phi_{l}\phi_{l}\right)_{\mathbf{3}}\right)_{\mathbf{1}}e^cH_d
+\frac{y_{e4}}{\Lambda^3}\left(\left(L\phi_{l}\right)_{\mathbf{3}^{\prime}}\left(\phi_{l}\phi_{l}\right)_{\mathbf{3}^{\prime}}\right)_{\mathbf{1}}e^cH_d\,,
\end{eqnarray}
where all the couplings are real because of the CP symmetry. After the electroweak and $S_4$ flavour symmetry breaking by the VEV shown in Eq.~\eqref{eq:vev_phil}, one can obtain that the charged lepton mass matrix is diagonal with the masses
\begin{equation}
m_e=\left|\left(y_{e_2}-2y_{e_4}\right)\frac{v^3_{\phi_{l}}}{\Lambda^3}\right|v_d, \qquad
m_{\mu}=\left|2y_{\mu_1}\frac{v^2_{\phi_{l}}}{\Lambda^2}\right|v_d,\qquad m_{\tau}=\left|y_{\tau}\frac{v_{\phi_{l}}}{\Lambda}\right|v_d\,,
\end{equation}
where $v_{d}=\langle H_{d}\rangle$ is the VEV of the electroweak Higgs field $H_d$. Since the charged lepton mass matrix is diagonal, the hermitian combination $m^\dagger_lm_l$ is invariant under the action of the subgroup $Z^T_3$, i.e. $\rho^\dagger_{\mathbf{3}}(T)m^\dagger_lm_l\rho_{\mathbf{3}}(T)=m^\dagger_lm_l$. With the assignment in table~\ref{tab:tran_pro}, the tau, muon and electron masses arise at the one-flavon, two-flavons and three-flavons level respectively in our model. Consequently the charged lepton mass hierarchies are naturally reproduced
\begin{equation}
m_e:m_{\mu}:m_{\tau}\simeq\lambda^4_C:\lambda^2_C:1\,.
\end{equation}
We find that the subleading order corrections to the charged lepton
mass matrix will break the residual symmetry $Z^T_3$ but they are suppressed by $\lambda^2_C$ with respect to LO results, thus can be safely neglected.

In the neutrino sector, the lowest dimensional operators responsible for neutrino masses are
\begin{equation}
\label{eq:w_nu}w_{\nu}=\frac{y_{a}}{\Lambda}\left(L\phi_{a}\right)_{\mathbf{1}}H_{u}\nu^{c}_{\text{atm}}+
\frac{y_{s}}{\Lambda}\left(L\phi_{s}\right)_{\mathbf{1^\prime}}H_{u}\nu^{c}_{\text{sol}}
+\frac{x_{a}}{2}\nu^{c}_{\text{atm}}\nu^{c}_{\text{atm}}\xi_{a}
+\frac{x_{s}}{2}\nu^{c}_{\text{sol}}\nu^{c}_{\text{sol}}\xi_{s}\,,
\end{equation}
where the coupling constants $y_{a}$, $y_{s}$, $x_{a}$ and $x_{s}$ are restricted to be real by the imposed CP symmetry. Inserting the vacuum alignments in Eqs.~(\ref{eq:vev_atm}, \ref{eq:vev_phis}), we can read out the neutrino Dirac and Majorana mass matrices as follow,
\begin{equation}\label{eq:Nu_Mass}
m_{D}=\begin{pmatrix}
0 &~ - y_{a}v_{\phi_{a}} &~  y_{a}v_{\phi_{a}}   \\
y_{s}v_{\phi_{s}}  &~ \frac{5}{2}y_{s}v_{\phi_{s}} &~ -\frac{1}{2}y_{s}v_{\phi_{s}}
\end{pmatrix}\frac{v_{u}}{\Lambda},
\qquad\quad  m_{N}=\begin{pmatrix}
x_{a}v_{\xi_{a}}  &  0  \\
0  &  x_{s}v_{\xi_{s}}
\end{pmatrix}\,,
\end{equation}
with $v_{u}=\langle H_{u}\rangle$. Applying the seesaw formula, we obtain the light neutrino mass matrix $m_{\nu}$ in Eq.~\eqref{eq:mnu_model} with
\begin{equation}
m_a=\left|\frac{y^2_{a}v^2_{\phi_{a}}}{x_{a}v_{\xi_{a}}}\frac{v^2_{u}}{\Lambda^2}\right|,\quad
m_s=\left|\frac{y^2_{s}v^2_{\phi_{s}}}{x_{s}v_{\xi_{s}}}\frac{v^2_{u}}{\Lambda^2}\right|\,.
\end{equation}
Note that we have used the result shown in Eq.~\eqref{eq:vev_ratio_arg} under the assumption of $x_ax_s>0$. In the case of $x_ax_s<0$, in order to obtain the desired value $\eta=-\pi/2$, the phase of the ratio $\frac{v^2_{\phi_{s}}v_{\xi_{a}}}{v^2_{\phi_{a}}v_{\xi_{s}}}$ should be $\pi/2$, i.e. we could choose the right side of Eq.~\eqref{eq:vev_ratio_arg} as $\pi/2$. In short, the neutrino mass matrix of the new Littlest Seesaw model is realized exactly, hence the phenomenological predictions in section~\ref{sec:spec_mix} follows immediately.
The NLO contributions to the neutrino mass matrices in Eq.~\eqref{eq:Nu_Mass} are found to be suppressed by $\lambda^2_C$ and consequently we will not discuss them.

Similar to other discrete flavour symmetry models, the solution of the vacuum alignment problem requires complicated constructions in our model and some new superfields which are SM singlets are introduced, as shown above.
Recently it was suggested that the complexity of the vacuum alignment problem can be reduced if modular invariance plays the role of flavour symmetry~\cite{Feruglio:2017spp}. In particular, we find that CSD($n$) model with $n=1+\sqrt{6}$ can be naturally obtained if the VEV of the complex modulus $\tau$ is at certain fixed point~\cite{Gui-JunDing:2019wap}. We expect that the desired alignment corresponding to $x=-1/2$ can also be reproduced from some modular group, such that our model could be simplified considerably.

\subsection{Charged lepton  flavour violating radiative decays}

In the following, we shall present the predictions for charged lepton flavour violating (LFV) radiative decays. It is usually assumed that the SUSY breaking mechanism is flavour blind at some high energy scale. In the minimal supergravity scenario, the slepton mass matrices are diagonal and universal in flavour and the trilinear couplings are proportional to the Yukawa couplings at the GUT scale. Non-vanishing off-diagonal elements are generated in both the slepton mass matrices and the trilinear couplings because of the renormalization group running effect at low energy, 
leading to charged lepton flavour violation processes induced in SUSY models. In the mass insertion and leading log approximations, the branching ratio of the charged lepton LFV radiative decay is given to good approximation by~\cite{Borzumati:1986qx,Hisano:1995cp}
\begin{equation}\label{eq:LFV_rad}
Br(l_i\to l_j\gamma)\simeq\frac{\alpha^3}{G^2_Fm^8_s}Br(l_i\to l_j\bar{\nu}_{j}\nu_{i})|(m^2_{\tilde{L}})_{ij}|^2\tan^2\beta\,,
\end{equation}
where $G_F$ is the Fermi coupling constant and $m_s$ is the characteristic mass scale of the SUSY particle in the loop with
\begin{equation}
m^8_s\simeq0.5m^2_0M^2_{1/2}(m^2_0+0.6M^2_{1/2})^2\,.
\end{equation}
The slepton doublet mass squared $m^2_{\tilde{L}}$ arises from the renormalization group evolution. To an excellent approximation, the renormalization group result has the form
\begin{equation}
(m^2_{\tilde{L}})_{i\neq j}\simeq -\frac{1}{8\pi^2}(3m^2_0+A^2_0)(\lambda^\dagger L \lambda)_{ij}\,,
\end{equation}
where $\lambda$ is the neutrino Yukawa coupling matrix given in Eq.~\eqref{eq:nu_Yukawa} and the factor $L$ is defined as
\begin{equation}
L=\text{diag}\left(\log\frac{M_G}{M_1},\log\frac{M_G}{M_2}\right)\,,
\end{equation}
with the GUT scale $M_G\simeq 2\times 10^{16}$ GeV. For our model, we find the expressions of $(m^2_{\tilde{L}})_{i\neq j}$ are as follows,
\begin{eqnarray}
\nonumber  && (m^2_{\tilde{L}})_{21}\simeq -\frac{5 |b|^2 \left(A^2_0+3 m^2_0\right) }{16 \pi ^2}\log \left(\frac{M_G}{M_2}\right)\,, \\
\nonumber  && (m^2_{\tilde{L}})_{31}\simeq \frac{|b|^2 \left(A^2_0+3 m^2_0\right) }{16 \pi ^2}\log \left(\frac{M_G}{M_2}\right)\,, \\
 && (m^2_{\tilde{L}})_{32}\simeq \frac{A^2_0+3 m^2_0}{32 \pi ^2}\left[4 |a|^2 \log \left(\frac{M_G}{M_1}\right)+5 |b|^2\log \left(\frac{M_G}{M_2}\right)\right]\,.
\end{eqnarray}
As shown in Eq.~\eqref{eq:MSSM_M1}, the right-handed neutrino mass $M_1$ is fixed to be $3.992\times 10^{10}$ GeV by leptogenesis. The best fit value of $r\equiv m_s/m_a$ is $0.145$ in our model, consequently it is natural to take $M_2\simeq3\times 10^{11}$ GeV. The measured values of the lepton mixing angles and neutrino masses fix $m_s=|b|^2v^2_u/M_2=3.243$ meV which leads to $b\simeq5.783\times 10^{-3}$. For typical values of the soft SUSY  breaking parameters $m_0=140\,\text{GeV}$, $M_{1/2}=600\, \text{GeV}$, $A_0=0$ and $\tan\beta=5$, we find the branching ratios of the charged lepton flavour violating radiative decays to be 
\begin{equation}
Br(\mu\to e\gamma)\simeq 1.745\times 10^{-16}\,, \quad
Br(\tau\to e\gamma)\simeq 1.244\times 10^{-18}\,, \quad
Br(\tau\to \mu\gamma)\simeq 2.647\times 10^{-17}\,,
\end{equation}
which are safely below the present experimental upper limits~\cite{Tanabashi:2018oca}. 

\subsection{UV completion}

In our model, we see that all interactions are renormalizable except the
the charged lepton Yukawa couplings in Eq.~\eqref{eq:ch_Yukawa} and the neutrino Yukawa couplings in Eq.~\eqref{eq:w_nu}. In the following, we shall give a UV completion which gives rise to these non-renormalizable operators upon integrating the heavy messengers fields. In order to generate the high dimensional operators relevant for charged lepton masses in Eq.~\eqref{eq:ch_Yukawa}, we introduce three pairs of messenger fields $\Sigma_i$ and $\Sigma^c_i$ with $i=1, 2, 3$ which transform under the flavour symmetry $S_4\times Z_5\times Z_8$ as follows
\begin{eqnarray}
\nonumber&&\Sigma_1\sim (\mathbf{3}, 1, 1),~~~~~~~\Sigma^c_1\sim (\mathbf{3}, 1, 1)\,,\\
\nonumber&&\Sigma_2\sim (\mathbf{3}', \omega_5, \omega^7_8),~~~~~\Sigma^c_2\sim (\mathbf{3}',\omega^4_5, \omega_8)\,,\\
&&\Sigma_3\sim (\mathbf{3}, \omega^2_5, -i),~~~~\Sigma^c_3\sim (\mathbf{3},  \omega^3_5, i)\,.
\end{eqnarray}
The chiral superfields $\Sigma_i$ and $\Sigma^c_i$ are singlets under the standard model gauge group and they carry hypercharges $Y=-1$ and $Y=1$ respectively, and their $U(1)_R$ charges are all $+1$. The renormalizable terms containing these messenger fields read as
\begin{eqnarray}
\nonumber w^{\mathrm{UV}}_l&=&g_1\left(L\Sigma^c_1\right)_{\mathbf{1}}H_d+g_2\left(\Sigma_1\phi_l\right)_{\mathbf{1}}\tau^c+g_3\left(\left(\Sigma_1\Sigma^c_2\right)_{\mathbf{3}}\phi_l\right)_{\mathbf{1}}
+g_4\left(\Sigma_2\phi_l\right)_{\mathbf{1}'}\mu^c+g_5\left(\left(\Sigma_2\Sigma^c_3\right)_{\mathbf{3}}\phi_l\right)_{\mathbf{1}}\\
&&+g_6\left(\Sigma_3\phi_l\right)_{\mathbf{1}}e^c+M_{\Sigma_1}\left(\Sigma_1\Sigma^c_1\right)_{\mathbf{1}}
+M_{\Sigma_2}\left(\Sigma_2\Sigma^c_2\right)_{\mathbf{1}}+M_{\Sigma_3}\left(\Sigma_3\Sigma^c_3\right)_{\mathbf{1}}\,,
\end{eqnarray}
where all the couplings $g_{1, 2, 3, 4, 5, 6}$ and masses $M_{\Sigma_{1,2,3}}$ are fixed to be real by the CP symmetry. Integrating out the heavy messenger fields, we obtain the desired higher-dimensional operators in the effective theory,
\begin{equation}
w^{\mathrm{eff}}_l=-\frac{g_1g_2}{M_{\Sigma_1}}\left(L\phi_{l}\right)_{\mathbf{1}}\tau^{c}H_d+\frac{g_1g_3g_4}{M_{\Sigma_1}M_{\Sigma_2}}\left(L\left(\phi_{l}\phi_{l}\right)_{\mathbf{3^\prime}}\right)_{\mathbf{1^\prime}}\mu^{c}H_d
-\frac{g_1g_3g_5g_6}{M_{\Sigma_1}M_{\Sigma_2}M_{\Sigma_3}}\left(\left(L\phi_{l}\right)_{\mathbf{3}^{\prime}}\left(\phi_{l}\phi_{l}\right)_{\mathbf{3}^{\prime}}\right)_{\mathbf{1}}e^cH_d\,,
\end{equation}
which leads to a diagonal and hierarchical charged lepton mass matrix for the alignment of $\phi_l$ in Eq.~\eqref{eq:vev_phil}. The non-renormalizable neutrino Dirac couplings in Eq.~\eqref{eq:w_nu} can be generated with the help of the following heavy fields
\begin{eqnarray}
\nonumber&&\Sigma_{\nu_1}\sim(\mathbf{1}, 1, -i),~~~~\Sigma^c_{\nu_1}\sim(\mathbf{1},  1, i)\,,\\
&&\Sigma_{\nu_2}\sim(\mathbf{1}', 1, \omega^3_8),~~~~\Sigma^c_{\nu_2}\sim(\mathbf{1}', 1, \omega^5_8)\,,
\end{eqnarray}
which are all standard model doublets with hypercharge $Y=\pm\frac{1}{2}$ ($-$ for $\Sigma_{\nu_i}$ and $+$ for $\Sigma^c_{\nu_i}$). The relevant terms in the UV completion are given by
\begin{eqnarray}
\nonumber w^{\mathrm{UV}}_{\nu}&=&k_1\left(L\phi_a\right)_{\mathbf{1}}\Sigma^c_{\nu_1}
+k_2\Sigma_{\nu_1}\nu^c_{\text{atm}}H_u+k_3\left(L\phi_s\right)_{\mathbf{1}'}\Sigma^c_{\nu_2}
+k_4\Sigma_{\nu_2}\nu^c_{\text{sol}}H_u\\
&&+M_{\Sigma_{\nu_1}}\Sigma_{\nu_1}\Sigma^c_{\nu_1}+M_{\Sigma_{\nu_2}}\Sigma_{\nu_2}\Sigma^c_{\nu_2}\,,
\end{eqnarray}
where CP invariance requires the parameters $k_{1, 2, 3, 4}$ and $M_{\Sigma_{\nu_{1, 2}}}$ are real. Integrating out $\Sigma_{\nu_{1, 2}}$ and $\Sigma^c_{\nu_{1, 2}}$, we reproduce the desired terms
\begin{equation}
w^{\mathrm{eff}}_{\nu}=-\frac{k_1k_2}{M_{\Sigma_{\nu_1}}}\left(L\phi_a\right)_{\mathbf{1}}H_{u}\nu^{c}_{\text{atm}}
-\frac{k_3k_4}{M_{\Sigma_{\nu_2}}}\left(L\phi_{s}\right)_{\mathbf{1^\prime}}H_{u}\nu^{c}_{\text{sol}}\,.
\end{equation}

\section{\label{sec:Conclusion}Conclusion}

In this paper we have proposed and discussed a new Littlest Seesaw model, realized in the tri-direct CP approach,
in which the couplings of the two right-handed neutrinos to the lepton doublets are proportional to $(0,-1,1)$ and $(1,5/2,-1/2)$ respectively with the relative phase $\eta=-\pi/2$. We have shown that this model can give an excellent description of lepton flavour mixing, including an atmospheric neutrino mixing angle in the second octant,
in terms of only two input parameters. We also showed that the observed baryon asymmetry can be generated for the lightest right-handed neutrino mass $M_{1}=1.176\times 10^{11}$ GeV in SM and $M_{1}=3.992\times 10^{10}$ GeV in MSSM with $\tan\beta=5$. The model is based on the flavour symmetry $S_4\times Z_5\times Z_8$ in which the desired alignments and the phase $\eta=-\pi/2$ are achieved.

We emphasise that the model independent tri-direct CP approach is a quite predictive scheme for constructing neutrino mass models based on discrete flavour symmetry and CP symmetry, even without specialising
to a particular choice of the two real input parameters
$\eta$ and $x$. Here we have focussed on the $\mathcal{N}_1$ case where
the flavour symmetry $S_{4}$ and CP are broken to $Z^T_3$ in the charged lepton sector, $Z^{U}_2\times H^{\text{atm}}_{CP}$ in the atmospheric sector and $Z_2^{SU}\times H^{\text{sol}}_{CP}$ in the solar neutrino sector with  $H^{\text{atm}}_{CP}=\{1,U\}$ and $H^{\text{sol}}_{CP}=\{1,SU\}$, the vacuum alignment of $\phi_{\textrm{atm}}$ and $\phi_{\textrm{sol}}$ would be fixed to $\langle\phi_{\text{atm}}\rangle\propto\left(0,1, -1\right)^T$ and $\langle\phi_{\text{sol}}\rangle\propto\left(1, x, 2-x\right)^T$,
where importantly $x$ is {\em real} due to the residual CP symmetry. As a consequence, the lepton mixing matrix is determined to be the TM1 pattern, and the experimental data on neutrino mixing can be described very well.
Thus the structure is enforced by residual symmetry in tri-direct CP approach, with $S_4$ flavour symmetry yielding good agreement with the present data for many examples,
which include both the original Littlest Seesaw model and the new Littlest Seesaw model~\cite{Ding:2018fyz,Ding:2018tuj}.

It is interesting to compare the new Littlest Seesaw with $(x, \eta)=(-1/2, -\pi/2)$ to the original Littlest Seesaw model with $(x, \eta)=(3, 2\pi/3)$, $(-1, -2\pi/3)$~\cite{King:2013xba,King:2015dvf,Ballett:2016yod}, which also provides a good fit to the data, as summarized in table~\ref{tab:bf}. However we find that the new Littlest Seesaw with arguably simpler values $x=-1/2$, $\eta=-\pi/2$, can provide a better description to the experimental data than the original Littlest Seesaw. The mixing parameters are predicted to lie in quite narrow regions, and they are all within the reach of future neutrino experiments. The denominator of the phase $\eta=-\pi/2$ is the smallest one among the different benchmark values in table~\ref{tab:bf}, consequently the case of $x=-1/2$, $\eta=-\pi/2$ might be expected to be easier to realize in a concrete model than the original Littlest Seesaw and other cases listed in table~\ref{tab:bf}.

We emphasise that the choice $x=-1/2$ and $\eta=-\pi/2$ of the new Littlest Seesaw model, is both simpler and more successful than the original Littlest Seesaw model. As usual, all three lepton mixing angles, leptonic CP violation phases and three neutrino masses ($m_1=0$) only depend on two input parameters $m_a$ and $r=m_s/m_a$ whose values can be determined by the precisely measured neutrino mass squared differences $\Delta m^2_{21}$ and $\Delta m^2_{31}$. The comprehensive numerical analysis shows that all lepton mixing parameters and neutrino masses are restricted in rather narrow regions, as shown in Eq.~\eqref{eq:model_prediction}. The new Littlest Seesaw differs most markedly in its predictions for $\theta_{23}$ and $\delta_{CP}$. While the atmospheric mixing angle $\theta_{23}$ is predicted to be close to maximal in the original Littlest Seesaw model, it is predicted to be in the second octant and close to the current central value~\cite{Esteban:2018azc} in the new Littlest Seesaw model. Furthermore, we have extended the new Littlest Seesaw to 3RHN models in the section 2.2. In the 3RHN model, we obtain a smaller $\chi^2_{\text{min}}$ than the new Littlest Seesaw, and we find that the 2RHN model is a good approximation of the 3RHN model. Therefore our new Littlest Seesaw with 2RHN can be regarded as a decoupling limit of the 3RHN model.

The ``maximal'' phase $\eta=-\pi/2$ is the unique source of CP violation in the new Littlest Seesaw model, as usual controlling both low energy CP violation and the CP asymmetry in leptogenesis. Hence the CP violation which may be observed in neutrino oscillations is related to the baryon asymmetry of the Universe. We have studied the generation of the baryon asymmetry of the Universe through leptogenesis in the new Littlest Seesaw model. We have numerically solved the flavoured Boltzmann equations for the lepton asymmetries, and found that the observed excess of matter over antimatter can be produced for the lightest right-handed neutrino mass $M_{1}=1.176\times 10^{11}$ GeV in SM and $M_{1}=3.992\times 10^{10}$ GeV in MSSM with $\tan\beta=5$. We conclude that the new Littlest Seesaw model can give an excellent fit to the neutrino oscillation data and leptogenesis simultaneously.

Finally we have constructed a fully working explicit model based on the flavour group $S_4$ and CP symmetry
which fixes the values of $x=-1/2$ and $\eta=-\pi/2$ in
the new Littlest Seesaw model. The charged lepton mass hierarchy is naturally realized in our model, and the required vacua $\langle\phi_{a}\rangle\propto\left(0, 1, -1\right)^{T}$, $\langle\phi_{s}\rangle\propto\left(1,-1/2,5/2\right)^{T}$ and the relative phase $\eta=-\pi/2$ are readily generated through the supersymmetric $F$-term alignment mechanism. Furthermore, we have studied the predictions for the charged lepton radiative decays $\mu\rightarrow e\gamma$, $\tau\rightarrow e\gamma$ and $\tau\rightarrow\mu\gamma$, and have found that the resulting branch ratios are below the current experimental upper bounds. We have also presented a UV completion which gives rise to the non-renormalizable operators upon integrating out the heavy messenger fields.

It would be interesting to extend this predictive new Littlest Seesaw model to the quark sector to give a unified description of quark and lepton flavour mixing, for instance in the framework of a supersymmetric grand unified theory. We expect that the quark mass matrices would be related to the construction of the new Littlest Seesaw model in the lepton sector. This is left for future work.

\subsection*{Acknowledgements}

P.-T.\,C. and G.-J.\,D. acknowledge the support of the National Natural Science Foundation of China under Grant Nos 11522546 and 11835013.
S.\,F.\,K. acknowledges the STFC Consolidated Grant ST/L000296/1
and the European Union's Horizon 2020 research and innovation programme under the Marie Sk\l{}odowska-Curie grant agreements
Elusives ITN No.\ 674896 and InvisiblesPlus RISE No.\ 690575. C.-C.\, L. is supported by National Natural Science Foundation of China under Grant No 11847228, China Postdoctoral Science Foundation  Grant Nos. 2017M620258 and 2018T110617, CPSF-CAS Joint Foundation for Excellent Postdoctoral Fellows No. 2017LH0003,   the Fundamental Research Funds for the Central Universities under Grant No. WK2030040090, the Anhui Province Natural Science Foundation Grant No. 1908085QA24 and the CAS Center for Excellence in Particle Physics (CCEPP).

\section*{Appendix}

\setcounter{equation}{0}
\renewcommand{\theequation}{\thesection.\arabic{equation}}

\begin{appendix}

\section{\label{sec:S4_group}Group Theory of $S_{4}$}

In the present work, we adopt the same convention for the $S_4$ flavour symmetry group as~\cite{Ding:2013hpa,Li:2013jya}. The $S_4$ group is generated by three generators $S$, $T$ and $U$ which obey the relations
\begin{eqnarray}
S^2=T^3=U^2=(ST)^3=(SU)^2=(TU)^2=(STU)^4=1\,.
\end{eqnarray}
The group $S_4$ has $24$ elements and five irreducible representations: $\mathbf{1}$, $\mathbf{1^{\prime}}$, $\mathbf{2}$, $\mathbf{3}$ and $\mathbf{3^{\prime}}$. The representation matrices of the three generators in different irreducible representations are chosen to be the following form
\begin{equation}\label{tab:S4_rep}
\begin{array}{clll}
\mathbf{1},~\mathbf{1}^{'}:~ &\quad S=1,&\quad T=1, &\quad U=\pm1\,,\\
\mathbf{2}^{'}~:~ &\qquad S=\begin{pmatrix}
1 ~&~0 \\
0 ~&~1
\end{pmatrix},&\quad T=\begin{pmatrix}
\omega ~&~0 \\
0 ~&~\omega^2
\end{pmatrix},&\quad U=\begin{pmatrix}
0 ~&~ 1 \\
1 ~&~ 0
\end{pmatrix}\,, \\
\mathbf{3},~\mathbf{3}^{'}:~&\quad S=\frac{1}{3}\begin{pmatrix}
-1 ~&~ 2  ~&~ 2  \\
2  ~&~ -1  ~&~ 2 \\
2  ~&~ 2 ~&~ -1
\end{pmatrix},&\quad T=\begin{pmatrix}
1  ~&~  0 ~&~ 0 \\
0  ~&~  \omega^2  ~&~ 0 \\
0  ~&~  0   ~&~  \omega
\end{pmatrix},&\quad U=\mp\begin{pmatrix}
1 ~&~  0  ~&~  0  \\
0  ~&~ 0  ~&~ 1 \\
0 ~&~ 1  ~&~  0
\end{pmatrix}\,,
\end{array}
\end{equation}
with $\omega=e^{2\pi i/3}$. As has been shown in~\cite{Ding:2013hpa,Li:2013jya}, the  generalized CP transformation compatible with the $S_{4}$ flavour symmetry is of the same form as the flavour symmetry transformation in our working basis,
\begin{equation}\label{eq:S4_GCP}
X_{\bf{r}}=\rho_{\mathbf{r}}(g), \qquad g\in S_{4}\,,
\end{equation}
where $g$ can be any of the 24 group elements of $S_4$. The $S_4$ Clebsch-Gordan coefficients are frequently used when building a model based on $S_4$ flavour symmetry. We summarise the Kronecker products and Clebsch-Gordan coefficients in our basis in table~\ref{tab:S4_CG}.

\begin{table}[t!]
\centering
\begin{tabular}{|c| c| c|c| c | c| c| c| c| c| c |c | c| c| c|}  \hline \hline
\multicolumn{2}{|c|}{~~$\mathbf{1^\prime}\otimes\mathbf{2}=\mathbf{2}$~~} &\multicolumn{2}{|c|}{~ ~$\mathbf{1^\prime}\otimes\mathbf{3}=\mathbf{3^\prime}$~ ~}&\multicolumn{2}{|c|}{~  ~$\mathbf{1^\prime}\otimes\mathbf{3^\prime}=\mathbf{3}$} \\ \hline
 \multicolumn{2}{|c|}{} & \multicolumn{2}{|c|}{} & \multicolumn{2}{|c|}{} \\[-0.18in]

\multicolumn{2}{|c|}{~~$\mathbf{2}\sim
\left(\begin{array}{c}a b_1 \\-a b_2\end{array}\right)$
~ ~ }&\multicolumn{2}{|c|}{~
~  $\mathbf{3^\prime} \sim
\left(\begin{array}{c}a b_1  \\ a b_2  \\ a b_3 \end{array}\right)$
~~ }&\multicolumn{2}{|c|}{~
~$\mathbf{3}\sim
\left(\begin{array}{c} a b_1  \\ a b_2  \\ a b_3 \end{array}\right)$~~}
 \\ \hline \hline

\multicolumn{2}{|c|}{ ~$\mathbf{2}\otimes\mathbf{2}=\mathbf{1}\oplus\mathbf{1^\prime}\oplus\mathbf{2}$~} &\multicolumn{2}{|c|}{ ~$\mathbf{2}\otimes\mathbf{3}=\mathbf{3}\oplus\mathbf{3^\prime}$~} &\multicolumn{2}{|c|}{  ~$\mathbf{2}\otimes\mathbf{3^\prime}=\mathbf{3}\oplus\mathbf{3^\prime}$} \\ \hline
\multicolumn{2}{|c|}{} & \multicolumn{2}{|c|}{} & \multicolumn{2}{|c|}{}\\[-0.18in]
\multicolumn{2}{|c|}{~~ ${\mathbf1}\sim a_1 b_2+ a_2 b_1$~~  }&\multicolumn{2}{|c|}{ ~~ }&\multicolumn{2}{|c|}{ ~~}  \\
\multicolumn{2}{|c|}{} & \multicolumn{2}{|c|}{} & \multicolumn{2}{|c|}{}\\[-0.37in]
\multicolumn{2}{|c|}{} & \multicolumn{2}{|c|}{~
~${\mathbf3}\sim
\left(\begin{array}{c}  a_1 b_2+ a_2 b_3 \\  a_1 b_3+ a_2 b_1  \\  a_1 b_1+ a_2 b_2 \end{array}\right)$
~~ }&\multicolumn{2}{|c|}{~
~$\mathbf{3}\sim
\left(\begin{array}{c}  a_1 b_2- a_2 b_3 \\  a_1 b_3- a_2 b_1  \\  a_1 b_1- a_2 b_2 \end{array}\right)$ ~~} \\
\multicolumn{2}{|c|}{} & \multicolumn{2}{|c|}{} & \multicolumn{2}{|c|}{}\\[-0.43in]
\multicolumn{2}{|c|}{~~${\mathbf 1^\prime}\sim a_1 b_2- a_2 b_1$
~ ~} & \multicolumn{2}{|c|}{} & \multicolumn{2}{|c|}{}  \\
\multicolumn{2}{|c|}{} & \multicolumn{2}{|c|}{} & \multicolumn{2}{|c|}{}\\[-0.10in]
\multicolumn{2}{|c|}{~~${\mathbf2}\sim
\left(\begin{array}{c} a_2 b_2  \\  a_1 b_1 \end{array}\right)$
~ ~ }&\multicolumn{2}{|c|}{~
~  ${\mathbf3^\prime}\sim
\left(\begin{array}{c} a_1 b_2- a_2 b_3 \\  a_1 b_3- a_2 b_1  \\  a_1 b_1- a_2 b_2 \end{array}\right)$
~~ }&\multicolumn{2}{|c|}{~
~$\mathbf{3^\prime}\sim
\left(\begin{array}{c} a_1 b_2+ a_2 b_3 \\  a_1 b_3+ a_2 b_1  \\  a_1 b_1+ a_2 b_2\end{array}\right)$ ~~}
\\ \hline \hline
\multicolumn{3}{|c|}{ ~ ~$ \mathbf{3}\otimes\mathbf{3}=\mathbf{3^\prime}\otimes\mathbf{3^\prime}=\mathbf{1}\oplus\mathbf{2}\oplus\mathbf{3}\oplus\mathbf{3^\prime}$~~ }&\multicolumn{3}{|c|}{ ~ ~$\mathbf{3}\otimes\mathbf{3^\prime}=\mathbf{1^\prime}\oplus\mathbf{2}\oplus\mathbf{3}\oplus\mathbf{3^\prime}$~~} \\ \hline
\multicolumn{3}{|c|}{} & \multicolumn{3}{|c|}{} \\[-0.16in]
 \multicolumn{3}{|c|}{~~ ${\mathbf1}\sim a_1 b_1+ a_2 b_3+ a_3 b_2$ ~~}&\multicolumn{3}{|c|}{ ~~ ${\mathbf1^\prime}\sim a_1 b_1+ a_2 b_3+ a_3 b_2$ ~~} \\
\multicolumn{3}{|c|}{} & \multicolumn{3}{|c|}{}\\[-0.16in]
\multicolumn{3}{|c|}{~~${\mathbf2}\sim
\left(\begin{array}{c} a_2 b_2+ a_1 b_3+ a_3 b_1  \\  a_3 b_3+ a_1 b_2+ a_2 b_1 \end{array}\right)$
~~ }&\multicolumn{3}{|c|}{ ~   ~ ${\mathbf2}\sim
\left(\begin{array}{c} a_2 b_2+ a_1 b_3+ a_3 b_1  \\ -( a_3 b_3+ a_1 b_2+ a_2 b_1) \end{array}\right)$~~}\\
\multicolumn{3}{|c|}{} & \multicolumn{3}{|c|}{}\\[-0.16in]
\multicolumn{3}{|c|}{~ ~${\mathbf3}\sim
\left(\begin{array}{c} a_2 b_3- a_3 b_2  \\  a_1 b_2- a_2 b_1  \\  a_3 b_1- a_1 b_3 \end{array}\right)$
~~ }&\multicolumn{3}{|c|}{~
~  ${\mathbf3}\sim
\left(\begin{array}{c}2 a_1 b_1-  a_2 b_3- a_3 b_2  \\ 2 a_3 b_3-  a_1 b_2- a_2 b_1  \\
 2 a_2 b_2- a_3 b_1- a_1 b_3 \end{array}\right)$
~~ } \\
\multicolumn{3}{|c|}{} & \multicolumn{3}{|c|}{} \\[-0.16in]
\multicolumn{3}{|c|}{~~ ${\mathbf3^\prime}\sim
\left(\begin{array}{c} 2 a_1 b_1- a_2 b_3- a_3 b_2  \\  2 a_3 b_3- a_1 b_2- a_2 b_1  \\
2 a_2 b_2- a_3 b_1- a_1 b_3 \end{array}\right)$
~~ }&\multicolumn{3}{|c|}{~
~${\mathbf3^\prime}\sim
\left(\begin{array}{c} a_2 b_3- a_3 b_2  \\  a_1 b_2- a_2 b_1  \\  a_3 b_1- a_1 b_3 \end{array}\right)$
~~ }\\ \hline\hline

\end{tabular}
\caption{\label{tab:S4_CG}The Kronecker products and Clebsch-Gordan coefficients of $S_4$ group~\cite{Ding:2013hpa,Li:2013jya}. We use $a_i$ to indicate the elements of the first representation of the product and $b_i$ to indicate those of the second representation.
}
\end{table}

\end{appendix}

\vskip 2cm

\vskip 2cm

\providecommand{\href}[2]{#2}\begingroup\raggedright\endgroup


\begin{thebibliography}{10}

\bibitem{King:2013eh}
S.~F. King and C.~Luhn, ``{Neutrino Mass and Mixing with Discrete Symmetry},''
  \href{http://dx.doi.org/10.1088/0034-4885/76/5/056201}{{\em Rept. Prog.
  Phys.} {\bfseries 76} (2013) 056201},
\href{http://arxiv.org/abs/1301.1340}{{\ttfamily arXiv:1301.1340 [hep-ph]}}.

\bibitem{King:2015aea}
S.~F. King, ``{Models of Neutrino Mass, Mixing and CP Violation},''
  \href{http://dx.doi.org/10.1088/0954-3899/42/12/123001}{{\em J. Phys.}
  {\bfseries G42} (2015) 123001},
\href{http://arxiv.org/abs/1510.02091}{{\ttfamily arXiv:1510.02091 [hep-ph]}}.

\bibitem{Minkowski:1977sc}
P.~Minkowski, ``{$\mu \to e\gamma$ at a Rate of One Out of $10^{9}$ Muon
  Decays?},''
\href{http://dx.doi.org/10.1016/0370-2693(77)90435-X}{{\em Phys. Lett.}
  {\bfseries 67B} (1977) 421--428}.

\bibitem{Mohapatra:1979ia}
R.~N. Mohapatra and G.~Senjanovic, ``{Neutrino Mass and Spontaneous Parity
  Nonconservation},'' \href{http://dx.doi.org/10.1103/PhysRevLett.44.912}{{\em
  Phys. Rev. Lett.} {\bfseries 44} (1980) 912}.
[,231(1979)].

\bibitem{Schechter:1980gr}
J.~Schechter and J.~W.~F. Valle, ``{Neutrino Masses in SU(2) x U(1)
  Theories},''
\href{http://dx.doi.org/10.1103/PhysRevD.22.2227}{{\em Phys. Rev.} {\bfseries
  D22} (1980) 2227}.

\bibitem{King:1998jw}
S.~F. King, ``{Atmospheric and solar neutrinos with a heavy singlet},''
  \href{http://dx.doi.org/10.1016/S0370-2693(98)01055-7}{{\em Phys. Lett.}
  {\bfseries B439} (1998) 350--356},
\href{http://arxiv.org/abs/hep-ph/9806440}{{\ttfamily arXiv:hep-ph/9806440
  [hep-ph]}}.

\bibitem{King:1999cm}
S.~F. King, ``{Atmospheric and solar neutrinos from single right-handed
  neutrino dominance and U(1) family symmetry},''
  \href{http://dx.doi.org/10.1016/S0550-3213(99)00542-8}{{\em Nucl. Phys.}
  {\bfseries B562} (1999) 57--77},
\href{http://arxiv.org/abs/hep-ph/9904210}{{\ttfamily arXiv:hep-ph/9904210
  [hep-ph]}}.

\bibitem{Esteban:2018azc}
I.~Esteban, M.~C. Gonzalez-Garcia, A.~Hernandez-Cabezudo, M.~Maltoni, and
  T.~Schwetz, ``{Global analysis of three-flavour neutrino oscillations:
  synergies and tensions in the determination of $\theta_{23}, \delta_{CP}$,
  and the mass ordering},''
  \href{http://dx.doi.org/10.1007/JHEP01(2019)106}{{\em JHEP} {\bfseries 01}
  (2019) 106},
\href{http://arxiv.org/abs/1811.05487}{{\ttfamily arXiv:1811.05487 [hep-ph]}}.

\bibitem{King:1999mb}
S.~F. King, ``{Large mixing angle MSW and atmospheric neutrinos from single
  right-handed neutrino dominance and U(1) family symmetry},''
  \href{http://dx.doi.org/10.1016/S0550-3213(00)00109-7}{{\em Nucl. Phys.}
  {\bfseries B576} (2000) 85--105},
\href{http://arxiv.org/abs/hep-ph/9912492}{{\ttfamily arXiv:hep-ph/9912492
  [hep-ph]}}.

\bibitem{Frampton:2002qc}
P.~H. Frampton, S.~L. Glashow, and T.~Yanagida, ``{Cosmological sign of
  neutrino CP violation},''
  \href{http://dx.doi.org/10.1016/S0370-2693(02)02853-8}{{\em Phys. Lett.}
  {\bfseries B548} (2002) 119--121},
\href{http://arxiv.org/abs/hep-ph/0208157}{{\ttfamily arXiv:hep-ph/0208157
  [hep-ph]}}.

\bibitem{King:2002nf}
S.~F. King, ``{Constructing the large mixing angle MNS matrix in seesaw models
  with right-handed neutrino dominance},''
  \href{http://dx.doi.org/10.1088/1126-6708/2002/09/011}{{\em JHEP} {\bfseries
  09} (2002) 011},
\href{http://arxiv.org/abs/hep-ph/0204360}{{\ttfamily arXiv:hep-ph/0204360
  [hep-ph]}}.

\bibitem{Guo:2006qa}
W.-l. Guo, Z.-z. Xing, and S.~Zhou, ``{Neutrino Masses, Lepton Flavor Mixing
  and Leptogenesis in the Minimal Seesaw Model},''
  \href{http://dx.doi.org/10.1142/S0218301307004898}{{\em Int. J. Mod. Phys.}
  {\bfseries E16} (2007) 1--50},
\href{http://arxiv.org/abs/hep-ph/0612033}{{\ttfamily arXiv:hep-ph/0612033
  [hep-ph]}}.

\bibitem{Harigaya:2012bw}
K.~Harigaya, M.~Ibe, and T.~T. Yanagida, ``{Seesaw Mechanism with Occam's
  Razor},'' \href{http://dx.doi.org/10.1103/PhysRevD.86.013002}{{\em Phys.
  Rev.} {\bfseries D86} (2012) 013002},
\href{http://arxiv.org/abs/1205.2198}{{\ttfamily arXiv:1205.2198 [hep-ph]}}.

\bibitem{Zhang:2015tea}
J.~Zhang and S.~Zhou, ``{A Further Study of the Frampton-Glashow-Yanagida Model
  for Neutrino Masses, Flavor Mixing and Baryon Number Asymmetry},''
  \href{http://dx.doi.org/10.1007/JHEP09(2015)065}{{\em JHEP} {\bfseries 09}
  (2015) 065},
\href{http://arxiv.org/abs/1505.04858}{{\ttfamily arXiv:1505.04858 [hep-ph]}}.

\bibitem{King:2005bj}
S.~F. King, ``{Predicting neutrino parameters from SO(3) family symmetry and
  quark-lepton unification},''
  \href{http://dx.doi.org/10.1088/1126-6708/2005/08/105}{{\em JHEP} {\bfseries
  08} (2005) 105},
\href{http://arxiv.org/abs/hep-ph/0506297}{{\ttfamily arXiv:hep-ph/0506297
  [hep-ph]}}.

\bibitem{Antusch:2011ic}
S.~Antusch, S.~F. King, C.~Luhn, and M.~Spinrath, ``{Trimaximal mixing with
  predicted $\theta_{13}$ from a new type of constrained sequential
  dominance},'' \href{http://dx.doi.org/10.1016/j.nuclphysb.2011.11.009}{{\em
  Nucl. Phys.} {\bfseries B856} (2012) 328--341},
\href{http://arxiv.org/abs/1108.4278}{{\ttfamily arXiv:1108.4278 [hep-ph]}}.

\bibitem{King:2013iva}
S.~F. King, ``{Minimal predictive see-saw model with normal neutrino mass
  hierarchy},'' \href{http://dx.doi.org/10.1007/JHEP07(2013)137}{{\em JHEP}
  {\bfseries 07} (2013) 137},
\href{http://arxiv.org/abs/1304.6264}{{\ttfamily arXiv:1304.6264 [hep-ph]}}.

\bibitem{King:2015dvf}
S.~F. King, ``{Littlest Seesaw},''
  \href{http://dx.doi.org/10.1007/JHEP02(2016)085}{{\em JHEP} {\bfseries 02}
  (2016) 085},
\href{http://arxiv.org/abs/1512.07531}{{\ttfamily arXiv:1512.07531 [hep-ph]}}.

\bibitem{King:2016yvg}
S.~F. King and C.~Luhn, ``{Littlest Seesaw model from S$_{4} \times$ U(1)},''
  \href{http://dx.doi.org/10.1007/JHEP09(2016)023}{{\em JHEP} {\bfseries 09}
  (2016) 023},
\href{http://arxiv.org/abs/1607.05276}{{\ttfamily arXiv:1607.05276 [hep-ph]}}.

\bibitem{Ballett:2016yod}
P.~Ballett, S.~F. King, S.~Pascoli, N.~W. Prouse, and T.~Wang, ``{Precision
  neutrino experiments vs the Littlest Seesaw},''
  \href{http://dx.doi.org/10.1007/JHEP03(2017)110}{{\em JHEP} {\bfseries 03}
  (2017) 110},
\href{http://arxiv.org/abs/1612.01999}{{\ttfamily arXiv:1612.01999 [hep-ph]}}.

\bibitem{King:2018fqh}
S.~F. King, S.~Molina~Sedgwick, and S.~J. Rowley, ``{Fitting high-energy
  Littlest Seesaw parameters using low-energy neutrino data and
  leptogenesis},'' \href{http://dx.doi.org/10.1007/JHEP10(2018)184}{{\em JHEP}
  {\bfseries 10} (2018) 184},
\href{http://arxiv.org/abs/1808.01005}{{\ttfamily arXiv:1808.01005 [hep-ph]}}.

\bibitem{King:2013xba}
S.~F. King, ``{Minimal see-saw model predicting best fit lepton mixing
  angles},'' \href{http://dx.doi.org/10.1016/j.physletb.2013.06.013}{{\em Phys.
  Lett.} {\bfseries B724} (2013) 92--98},
\href{http://arxiv.org/abs/1305.4846}{{\ttfamily arXiv:1305.4846 [hep-ph]}}.

\bibitem{King:2013hoa}
S.~F. King, ``{A model of quark and lepton mixing},''
  \href{http://dx.doi.org/10.1007/JHEP01(2014)119}{{\em JHEP} {\bfseries 01}
  (2014) 119},
\href{http://arxiv.org/abs/1311.3295}{{\ttfamily arXiv:1311.3295 [hep-ph]}}.

\bibitem{Bjorkeroth:2014vha}
F.~Bj$\ddot{\rm{o}}$rkeroth and S.~F. King, ``{Testing constrained sequential dominance
  models of neutrinos},''
  \href{http://dx.doi.org/10.1088/0954-3899/42/12/125002}{{\em J. Phys.}
  {\bfseries G42} no.~12, (2015) 125002},
\href{http://arxiv.org/abs/1412.6996}{{\ttfamily arXiv:1412.6996 [hep-ph]}}.

\bibitem{Feruglio:2012cw}
F.~Feruglio, C.~Hagedorn, and R.~Ziegler, ``{Lepton Mixing Parameters from
  Discrete and CP Symmetries},''
  \href{http://dx.doi.org/10.1007/JHEP07(2013)027}{{\em JHEP} {\bfseries 07}
  (2013) 027},
\href{http://arxiv.org/abs/1211.5560}{{\ttfamily arXiv:1211.5560 [hep-ph]}}.

\bibitem{Holthausen:2012dk}
M.~Holthausen, M.~Lindner, and M.~A. Schmidt, ``{CP and Discrete Flavour
  Symmetries},'' \href{http://dx.doi.org/10.1007/JHEP04(2013)122}{{\em JHEP}
  {\bfseries 04} (2013) 122},
\href{http://arxiv.org/abs/1211.6953}{{\ttfamily arXiv:1211.6953 [hep-ph]}}.

\bibitem{Ding:2013hpa}
G.-J. Ding, S.~F. King, C.~Luhn, and A.~J. Stuart, ``{Spontaneous CP violation
  from vacuum alignment in $S_4$ models of leptons},''
  \href{http://dx.doi.org/10.1007/JHEP05(2013)084}{{\em JHEP} {\bfseries 05}
  (2013) 084},
\href{http://arxiv.org/abs/1303.6180}{{\ttfamily arXiv:1303.6180 [hep-ph]}}.

\bibitem{Ding:2013bpa}
G.-J. Ding, S.~F. King, and A.~J. Stuart, ``{Generalised CP and $A_4$ Family
  Symmetry},'' \href{http://dx.doi.org/10.1007/JHEP12(2013)006}{{\em JHEP}
  {\bfseries 12} (2013) 006},
\href{http://arxiv.org/abs/1307.4212}{{\ttfamily arXiv:1307.4212 [hep-ph]}}.

\bibitem{Li:2013jya}
C.-C. Li and G.-J. Ding, ``{Generalised CP and trimaximal $TM_1$ lepton mixing
  in $S_4$ family symmetry},''
  \href{http://dx.doi.org/10.1016/j.nuclphysb.2014.02.002}{{\em Nucl. Phys.}
  {\bfseries B881} (2014) 206--232},
\href{http://arxiv.org/abs/1312.4401}{{\ttfamily arXiv:1312.4401 [hep-ph]}}.

\bibitem{Ding:2013nsa}
G.-J. Ding and Y.-L. Zhou, ``{Predicting lepton flavor mixing from $\Delta$(48)
  and generalized $CP$ symmetries},''
  \href{http://dx.doi.org/10.1088/1674-1137/39/2/021001}{{\em Chin. Phys.}
  {\bfseries C39} no.~2, (2015) 021001},
\href{http://arxiv.org/abs/1312.5222}{{\ttfamily arXiv:1312.5222 [hep-ph]}}.

\bibitem{Ding:2014ssa}
G.-J. Ding and S.~F. King, ``{Generalized $CP$ and $\Delta(96)$ family
  symmetry},'' \href{http://dx.doi.org/10.1103/PhysRevD.89.093020}{{\em Phys.
  Rev.} {\bfseries D89} no.~9, (2014) 093020},
\href{http://arxiv.org/abs/1403.5846}{{\ttfamily arXiv:1403.5846 [hep-ph]}}.

\bibitem{Ding:2014hva}
G.-J. Ding and Y.-L. Zhou, ``{Lepton mixing parameters from $\Delta(48)$ family
  symmetry and generalised CP},''
  \href{http://dx.doi.org/10.1007/JHEP06(2014)023}{{\em JHEP} {\bfseries 06}
  (2014) 023},
\href{http://arxiv.org/abs/1404.0592}{{\ttfamily arXiv:1404.0592 [hep-ph]}}.

\bibitem{Li:2014eia}
C.-C. Li and G.-J. Ding, ``{Deviation from bimaximal mixing and leptonic CP
  phases in S$_{4}$ family symmetry and generalized CP},''
  \href{http://dx.doi.org/10.1007/JHEP08(2015)017}{{\em JHEP} {\bfseries 08}
  (2015) 017},
\href{http://arxiv.org/abs/1408.0785}{{\ttfamily arXiv:1408.0785 [hep-ph]}}.

\bibitem{Ding:2014ora}
G.-J. Ding, S.~F. King, and T.~Neder, ``{Generalised CP and $\Delta(6n^2)$
  family symmetry in semi-direct models of leptons},''
  \href{http://dx.doi.org/10.1007/JHEP12(2014)007}{{\em JHEP} {\bfseries 12}
  (2014) 007},
\href{http://arxiv.org/abs/1409.8005}{{\ttfamily arXiv:1409.8005 [hep-ph]}}.

\bibitem{Chen:2014wxa}
P.~Chen, C.-C. Li, and G.-J. Ding, ``{Lepton Flavor Mixing and CP Symmetry},''
  \href{http://dx.doi.org/10.1103/PhysRevD.91.033003}{{\em Phys. Rev.}
  {\bfseries D91} (2015) 033003},
\href{http://arxiv.org/abs/1412.8352}{{\ttfamily arXiv:1412.8352 [hep-ph]}}.

\bibitem{Everett:2015oka}
L.~L. Everett, T.~Garon, and A.~J. Stuart, ``{A Bottom-Up Approach to Lepton
  Flavor and CP Symmetries},''
  \href{http://dx.doi.org/10.1007/JHEP04(2015)069}{{\em JHEP} {\bfseries 04}
  (2015) 069},
\href{http://arxiv.org/abs/1501.04336}{{\ttfamily arXiv:1501.04336 [hep-ph]}}.

\bibitem{Branco:2015hea}
G.~C. Branco, I.~de~Medeiros~Varzielas, and S.~F. King, ``{Invariant approach
  to CP in family symmetry models},''
  \href{http://dx.doi.org/10.1103/PhysRevD.92.036007}{{\em Phys. Rev.}
  {\bfseries D92} no.~3, (2015) 036007},
\href{http://arxiv.org/abs/1502.03105}{{\ttfamily arXiv:1502.03105 [hep-ph]}}.

\bibitem{Li:2015jxa}
C.-C. Li and G.-J. Ding, ``{Lepton Mixing in $A_5$ Family Symmetry and
  Generalized CP},'' \href{http://dx.doi.org/10.1007/JHEP05(2015)100}{{\em
  JHEP} {\bfseries 05} (2015) 100},
\href{http://arxiv.org/abs/1503.03711}{{\ttfamily arXiv:1503.03711 [hep-ph]}}.

\bibitem{DiIura:2015kfa}
A.~Di~Iura, C.~Hagedorn, and D.~Meloni, ``{Lepton mixing from the interplay of
  the alternating group A$_{5}$ and CP},''
  \href{http://dx.doi.org/10.1007/JHEP08(2015)037}{{\em JHEP} {\bfseries 08}
  (2015) 037},
\href{http://arxiv.org/abs/1503.04140}{{\ttfamily arXiv:1503.04140 [hep-ph]}}.

\bibitem{Ballett:2015wia}
P.~Ballett, S.~Pascoli, and J.~Turner, ``{Mixing angle and phase correlations
  from A5 with generalized CP and their prospects for discovery},''
  \href{http://dx.doi.org/10.1103/PhysRevD.92.093008}{{\em Phys. Rev.}
  {\bfseries D92} no.~9, (2015) 093008},
\href{http://arxiv.org/abs/1503.07543}{{\ttfamily arXiv:1503.07543 [hep-ph]}}.

\bibitem{Branco:2015gna}
G.~C. Branco, I.~de~Medeiros~Varzielas, and S.~F. King, ``{Invariant approach
  to $\mathcal {CP}$ in unbroken $\Delta(27)$},''
  \href{http://dx.doi.org/10.1016/j.nuclphysb.2015.07.024}{{\em Nucl. Phys.}
  {\bfseries B899} (2015) 14--36},
\href{http://arxiv.org/abs/1505.06165}{{\ttfamily arXiv:1505.06165 [hep-ph]}}.

\bibitem{Chen:2015nha}
P.~Chen, C.-Y. Yao, and G.-J. Ding, ``{Neutrino Mixing from CP Symmetry},''
  \href{http://dx.doi.org/10.1103/PhysRevD.92.073002}{{\em Phys. Rev.}
  {\bfseries D92} no.~7, (2015) 073002},
\href{http://arxiv.org/abs/1507.03419}{{\ttfamily arXiv:1507.03419 [hep-ph]}}.

\bibitem{Ding:2015rwa}
G.-J. Ding and S.~F. King, ``{Generalized CP and $\Delta (3n^2)$ Family
  Symmetry for Semi-Direct Predictions of the PMNS Matrix},''
  \href{http://dx.doi.org/10.1103/PhysRevD.93.025013}{{\em Phys. Rev.}
  {\bfseries D93} (2016) 025013},
\href{http://arxiv.org/abs/1510.03188}{{\ttfamily arXiv:1510.03188 [hep-ph]}}.

\bibitem{Chen:2015siy}
P.~Chen, G.-J. Ding, F.~Gonzalez-Canales, and J.~W.~F. Valle, ``{Generalized
  $\mu-\tau$ reflection symmetry and leptonic CP violation},''
  \href{http://dx.doi.org/10.1016/j.physletb.2015.12.069}{{\em Phys. Lett.}
  {\bfseries B753} (2016) 644--652},
\href{http://arxiv.org/abs/1512.01551}{{\ttfamily arXiv:1512.01551 [hep-ph]}}.

\bibitem{Li:2016ppt}
C.-C. Li, C.-Y. Yao, and G.-J. Ding, ``{Lepton Mixing Predictions from Infinite
  Group Series $D^{(1)}_{9n, 3n}$ with Generalized CP},''
  \href{http://dx.doi.org/10.1007/JHEP05(2016)007}{{\em JHEP} {\bfseries 05}
  (2016) 007},
\href{http://arxiv.org/abs/1601.06393}{{\ttfamily arXiv:1601.06393 [hep-ph]}}.

\bibitem{Chen:2016ptr}
P.~Chen, G.-J. Ding, and S.~F. King, ``{Leptogenesis and residual CP
  symmetry},'' \href{http://dx.doi.org/10.1007/JHEP03(2016)206}{{\em JHEP}
  {\bfseries 03} (2016) 206},
\href{http://arxiv.org/abs/1602.03873}{{\ttfamily arXiv:1602.03873 [hep-ph]}}.

\bibitem{Yao:2016zev}
C.-Y. Yao and G.-J. Ding, ``{CP Symmetry and Lepton Mixing from a Scan of
  Finite Discrete Groups},''
  \href{http://dx.doi.org/10.1103/PhysRevD.94.073006}{{\em Phys. Rev.}
  {\bfseries D94} no.~7, (2016) 073006},
\href{http://arxiv.org/abs/1606.05610}{{\ttfamily arXiv:1606.05610 [hep-ph]}}.

\bibitem{Li:2016nap}
C.-C. Li, J.-N. Lu, and G.-J. Ding, ``{$A_4$ and CP symmetry and a model with
  maximal CP violation},''
  \href{http://dx.doi.org/10.1016/j.nuclphysb.2016.09.005}{{\em Nucl. Phys.}
  {\bfseries B913} (2016) 110--131},
\href{http://arxiv.org/abs/1608.01860}{{\ttfamily arXiv:1608.01860 [hep-ph]}}.

\bibitem{Lu:2016jit}
J.-N. Lu and G.-J. Ding, ``{Alternative Schemes of Predicting Lepton Mixing
  Parameters from Discrete Flavor and CP Symmetry},''
  \href{http://dx.doi.org/10.1103/PhysRevD.95.015012}{{\em Phys. Rev.}
  {\bfseries D95} no.~1, (2017) 015012},
\href{http://arxiv.org/abs/1610.05682}{{\ttfamily arXiv:1610.05682 [hep-ph]}}.

\bibitem{Everett:2016jsk}
L.~L. Everett and A.~J. Stuart, ``{Lepton Sector Phases and Their Roles in
  Flavor and Generalized CP Symmetries},''
  \href{http://dx.doi.org/10.1103/PhysRevD.96.035030}{{\em Phys. Rev.}
  {\bfseries D96} no.~3, (2017) 035030},
\href{http://arxiv.org/abs/1611.03020}{{\ttfamily arXiv:1611.03020 [hep-ph]}}.

\bibitem{Li:2017zmk}
C.-C. Li and G.-J. Ding, ``{Implications of residual CP symmetry for
  leptogenesis in a model with two right-handed neutrinos},''
  \href{http://dx.doi.org/10.1103/PhysRevD.96.075005}{{\em Phys. Rev.}
  {\bfseries D96} no.~7, (2017) 075005},
\href{http://arxiv.org/abs/1701.08508}{{\ttfamily arXiv:1701.08508 [hep-ph]}}.

\bibitem{Li:2017abz}
C.-C. Li, J.-N. Lu, and G.-J. Ding, ``{Toward a unified interpretation of quark
  and lepton mixing from flavor and CP symmetries},''
  \href{http://dx.doi.org/10.1007/JHEP02(2018)038}{{\em JHEP} {\bfseries 02}
  (2018) 038},
\href{http://arxiv.org/abs/1706.04576}{{\ttfamily arXiv:1706.04576 [hep-ph]}}.

\bibitem{Lu:2018oxc}
J.-N. Lu and G.-J. Ding, ``{Quark and lepton mixing patterns from a common
  discrete flavor symmetry with a generalized CP symmetry},''
  \href{http://dx.doi.org/10.1103/PhysRevD.98.055011}{{\em Phys. Rev.}
  {\bfseries D98} no.~5, (2018) 055011},
\href{http://arxiv.org/abs/1806.02301}{{\ttfamily arXiv:1806.02301 [hep-ph]}}.

\bibitem{Lu:2019gqp}
J.-N. Lu and G.-J. Ding, ``{Dihedral flavor group as the key to understand
  quark and lepton flavor mixing},''
  \href{http://dx.doi.org/10.1007/JHEP03(2019)056}{{\em JHEP} {\bfseries 03}
  (2019) 056},
\href{http://arxiv.org/abs/1901.07414}{{\ttfamily arXiv:1901.07414 [hep-ph]}}.

\bibitem{Chen:2018lsv}
P.~Chen, S.~Centelles~Chuli¨¢, G.-J. Ding, R.~Srivastava, and J.~W.~F. Valle,
  ``{Neutrino Predictions from Generalized CP Symmetries of Charged Leptons},''
  \href{http://dx.doi.org/10.1007/JHEP07(2018)077}{{\em JHEP} {\bfseries 07}
  (2018) 077},
\href{http://arxiv.org/abs/1802.04275}{{\ttfamily arXiv:1802.04275 [hep-ph]}}.

\bibitem{Hagedorn:2016lva}
C.~Hagedorn and E.~Molinaro, ``{Flavor and CP symmetries for leptogenesis and $0\nu\beta\beta$ decay},''
  \href{http://dx.doi.org/10.1016/j.nuclphysb.2017.03.015}{{\em Nucl. Phys.}
  {\bfseries B919} (2017) 404--469},
\href{http://arxiv.org/abs/1602.04206}{{\ttfamily arXiv:1602.04206 [hep-ph]}}.

\bibitem{Delgadillo:2018tza}
L.~A. Delgadillo, L.~L. Everett, R.~Ramos, and A.~J. Stuart, ``{Predictions for
  the Dirac CP-Violating Phase from Sum Rules},''
  \href{http://dx.doi.org/10.1103/PhysRevD.97.095001}{{\em Phys. Rev.}
  {\bfseries D97} no.~9, (2018) 095001},
\href{http://arxiv.org/abs/1801.06377}{{\ttfamily arXiv:1801.06377 [hep-ph]}}.

\bibitem{Ding:2018fyz}
G.-J. Ding, S.~F. King, and C.-C. Li, ``{Tri-Direct CP in the Littlest Seesaw
  Playground},'' \href{http://dx.doi.org/10.1007/JHEP12(2018)003}{{\em JHEP}
  {\bfseries 12} (2018) 003},
\href{http://arxiv.org/abs/1807.07538}{{\ttfamily arXiv:1807.07538 [hep-ph]}}.

\bibitem{Ding:2018tuj}
G.-J. Ding, S.~F. King, and C.-C. Li, ``{Lepton Mixing Predictions from $S_4$
  in the Tri-Direct CP approach to Two Right-handed Neutrino Models},''
\href{http://arxiv.org/abs/1811.12340}{{\ttfamily arXiv:1811.12340 [hep-ph]}}.

\bibitem{Ding:2017hdv}
G.-J. Ding, S.~F. King, and C.-C. Li, ``{Golden Littlest Seesaw},''
  \href{http://dx.doi.org/10.1016/j.nuclphysb.2017.10.019}{{\em Nucl. Phys.}
  {\bfseries B925} (2017) 470--499},
\href{http://arxiv.org/abs/1705.05307}{{\ttfamily arXiv:1705.05307 [hep-ph]}}.

\bibitem{Jarlskog:1985ht}
C.~Jarlskog, ``{Commutator of the Quark Mass Matrices in the Standard
  Electroweak Model and a Measure of Maximal CP Violation},''
\href{http://dx.doi.org/10.1103/PhysRevLett.55.1039}{{\em Phys. Rev. Lett.}
  {\bfseries 55} (1985) 1039}.

\bibitem{Branco:1986gr}
G.~C. Branco, L.~Lavoura, and M.~N. Rebelo, ``{Majorana Neutrinos and {CP}
  Violation in the Leptonic Sector},''
\href{http://dx.doi.org/10.1016/0370-2693(86)90307-2}{{\em Phys. Lett.}
  {\bfseries B180} (1986) 264--268}.

\bibitem{Aghanim:2018eyx}
{\bfseries Planck} Collaboration, N.~Aghanim {\em et~al.}, ``{Planck 2018
  results. VI. Cosmological parameters},''
\href{http://arxiv.org/abs/1807.06209}{{\ttfamily arXiv:1807.06209
  [astro-ph.CO]}}.

\bibitem{Fukugita:1986hr}
M.~Fukugita and T.~Yanagida, ``{Baryogenesis Without Grand Unification},''
\href{http://dx.doi.org/10.1016/0370-2693(86)91126-3}{{\em Phys. Lett.}
  {\bfseries B174} (1986) 45--47}.

\bibitem{Nardi:2006fx}
E.~Nardi, Y.~Nir, E.~Roulet, and J.~Racker, ``{The Importance of flavor in
  leptogenesis},'' \href{http://dx.doi.org/10.1088/1126-6708/2006/01/164}{{\em
  JHEP} {\bfseries 01} (2006) 164},
\href{http://arxiv.org/abs/hep-ph/0601084}{{\ttfamily arXiv:hep-ph/0601084
  [hep-ph]}}.

\bibitem{Abada:2006ea}
A.~Abada, S.~Davidson, A.~Ibarra, F.~X. Josse-Michaux, M.~Losada, and
  A.~Riotto, ``{Flavour Matters in Leptogenesis},''
  \href{http://dx.doi.org/10.1088/1126-6708/2006/09/010}{{\em JHEP} {\bfseries
  09} (2006) 010},
\href{http://arxiv.org/abs/hep-ph/0605281}{{\ttfamily arXiv:hep-ph/0605281
  [hep-ph]}}.

\bibitem{Abada:2006fw}
A.~Abada, S.~Davidson, F.-X. Josse-Michaux, M.~Losada, and A.~Riotto, ``{Flavor
  issues in leptogenesis},''
  \href{http://dx.doi.org/10.1088/1475-7516/2006/04/004}{{\em JCAP} {\bfseries
  0604} (2006) 004},
\href{http://arxiv.org/abs/hep-ph/0601083}{{\ttfamily arXiv:hep-ph/0601083
  [hep-ph]}}.

\bibitem{Antusch:2006cw}
S.~Antusch, S.~F. King, and A.~Riotto, ``{Flavour-Dependent Leptogenesis with
  Sequential Dominance},''
  \href{http://dx.doi.org/10.1088/1475-7516/2006/11/011}{{\em JCAP} {\bfseries
  0611} (2006) 011},
\href{http://arxiv.org/abs/hep-ph/0609038}{{\ttfamily arXiv:hep-ph/0609038
  [hep-ph]}}.

\bibitem{Giudice:2003jh}
G.~F. Giudice, A.~Notari, M.~Raidal, A.~Riotto, and A.~Strumia, ``{Towards a
  complete theory of thermal leptogenesis in the SM and MSSM},''
  \href{http://dx.doi.org/10.1016/j.nuclphysb.2004.02.019}{{\em Nucl. Phys.}
  {\bfseries B685} (2004) 89--149},
\href{http://arxiv.org/abs/hep-ph/0310123}{{\ttfamily arXiv:hep-ph/0310123
  [hep-ph]}}.

\bibitem{Buchmuller:2004nz}
W.~Buchmuller, P.~Di~Bari, and M.~Plumacher, ``{Leptogenesis for
  pedestrians},'' \href{http://dx.doi.org/10.1016/j.aop.2004.02.003}{{\em
  Annals Phys.} {\bfseries 315} (2005) 305--351},
\href{http://arxiv.org/abs/hep-ph/0401240}{{\ttfamily arXiv:hep-ph/0401240
  [hep-ph]}}.

\bibitem{Covi:1996wh}
L.~Covi, E.~Roulet, and F.~Vissani, ``{CP violating decays in leptogenesis
  scenarios},'' \href{http://dx.doi.org/10.1016/0370-2693(96)00817-9}{{\em
  Phys. Lett.} {\bfseries B384} (1996) 169--174},
\href{http://arxiv.org/abs/hep-ph/9605319}{{\ttfamily arXiv:hep-ph/9605319
  [hep-ph]}}.

\bibitem{Buchmuller:2005eh}
W.~Buchmuller, R.~D. Peccei, and T.~Yanagida, ``{Leptogenesis as the origin of
  matter},''
  \href{http://dx.doi.org/10.1146/annurev.nucl.55.090704.151558}{{\em Ann. Rev.
  Nucl. Part. Sci.} {\bfseries 55} (2005) 311--355},
\href{http://arxiv.org/abs/hep-ph/0502169}{{\ttfamily arXiv:hep-ph/0502169
  [hep-ph]}}.

\bibitem{Davidson:2008bu}
S.~Davidson, E.~Nardi, and Y.~Nir, ``{Leptogenesis},''
  \href{http://dx.doi.org/10.1016/j.physrep.2008.06.002}{{\em Phys. Rept.}
  {\bfseries 466} (2008) 105--177},
\href{http://arxiv.org/abs/0802.2962}{{\ttfamily arXiv:0802.2962 [hep-ph]}}.

\bibitem{Harvey:1990qw}
J.~A. Harvey and M.~S. Turner, ``{Cosmological baryon and lepton number in the
  presence of electroweak fermion number violation},''
\href{http://dx.doi.org/10.1103/PhysRevD.42.3344}{{\em Phys. Rev.} {\bfseries
  D42} (1990) 3344--3349}.

\bibitem{Altarelli:2005yx}
G.~Altarelli and F.~Feruglio, ``{Tri-bimaximal neutrino mixing, A(4) and the
  modular symmetry},''
  \href{http://dx.doi.org/10.1016/j.nuclphysb.2006.02.015}{{\em Nucl. Phys.}
  {\bfseries B741} (2006) 215--235},
\href{http://arxiv.org/abs/hep-ph/0512103}{{\ttfamily arXiv:hep-ph/0512103
  [hep-ph]}}.

\bibitem{Feruglio:2017spp}
F.~Feruglio, \href{http://dx.doi.org/10.1142/9789813238053_0012}{``{Are
  neutrino masses modular forms?},''} in {\em From My Vast Repertoire ...:
  Guido Altarelli's Legacy}, A.~Levy, S.~Forte, and G.~Ridolfi, eds.,
  pp.~227--266.
\newblock 2019.
\newblock
\href{http://arxiv.org/abs/1706.08749}{{\ttfamily arXiv:1706.08749 [hep-ph]}}.
\newblock

\bibitem{Gui-JunDing:2019wap}
G.-J. Ding, S.~F. King, X.-G. Liu, and J.-N. Lu, ``{Modular S$_{4}$ and A$_{4}$
  symmetries and their fixed points: new predictive examples of lepton
  mixing},'' \href{http://dx.doi.org/10.1007/JHEP12(2019)030}{{\em JHEP}
  {\bfseries 12} (2019) 030},
\href{http://arxiv.org/abs/1910.03460}{{\ttfamily arXiv:1910.03460 [hep-ph]}}.

\bibitem{Borzumati:1986qx}
F.~Borzumati and A.~Masiero, ``{Large Muon and electron Number Violations in
  Supergravity Theories},''
\href{http://dx.doi.org/10.1103/PhysRevLett.57.961}{{\em Phys. Rev. Lett.}
  {\bfseries 57} (1986) 961}.

\bibitem{Hisano:1995cp}
J.~Hisano, T.~Moroi, K.~Tobe, and M.~Yamaguchi, ``{Lepton flavor violation via
  right-handed neutrino Yukawa couplings in supersymmetric standard model},''
  \href{http://dx.doi.org/10.1103/PhysRevD.53.2442}{{\em Phys. Rev.} {\bfseries
  D53} (1996) 2442--2459},
\href{http://arxiv.org/abs/hep-ph/9510309}{{\ttfamily arXiv:hep-ph/9510309
  [hep-ph]}}.

\bibitem{Tanabashi:2018oca}
{\bfseries Particle Data Group} Collaboration, M.~Tanabashi {\em et~al.},
  ``{Review of Particle Physics},''
\href{http://dx.doi.org/10.1103/PhysRevD.98.030001}{{\em Phys. Rev.} {\bfseries
  D98} no.~3, (2018) 030001}.

\end{thebibliography}
\end{document}